\documentclass[preprint,10pt]{article}
\usepackage{longtable}
\usepackage{amsmath}
\usepackage{enumerate}
\usepackage{array}
\usepackage{subfig}
\usepackage{graphicx}
\usepackage{float}
\usepackage{amsfonts}
\usepackage{epstopdf}
\usepackage{geometry}
\usepackage[cm]{fullpage}
\usepackage{multicol}
\usepackage{authblk}
\usepackage{url}
\usepackage{cite}

\usepackage[titletoc,title]{appendix}

\title{Separating temporal and topological effects in walk-based network centrality}
\author[1]{Ewan R. Colman}
\author[2]{Nathaniel Charlton}
\affil[1]{\small{Department of Biology, Georgetown University, Washington, DC 20057, U.S.A}}
\affil[2]{\small{CountingLab Ltd., Reading, United Kingdom, RG6 6AX}}

\begin{document}
\maketitle
\vspace{-0.7cm}
\begin{abstract}
\noindent
The recently introduced concept of dynamic communicability is a valuable tool for ranking the importance of nodes in a temporal network. Two metrics, broadcast score and receive score, were introduced to measure the centrality of a node with respect to a model of contagion based on time-respecting walks. This article examines the temporal and structural factors influencing these metrics by considering a versatile stochastic temporal network model. We analytically derive formulae to accurately predict the expectation of the broadcast and receive scores when one or more columns in a temporal edge-list are shuffled. These methods are then applied to two publicly available data-sets and we quantify how much the centrality of each individual depends on structural or temporal influences. From our analysis we highlight two practical contributions: a way to control for temporal variation when computing dynamic communicability, and the conclusion that the broadcast and receive scores can, under a range of circumstances, be replaced by the row and column sums of the matrix exponential of a weighted adjacency matrix given by the data.
\end{abstract}
\vspace{0.5cm}
\begin{multicols}{2}

\section{Introduction}
Epidemics, viral marketing, cultural diffusion, the distribution of food in ant colonies, and the flow of information within the human brain, are amongst a growing number of applications of network theory which currently reside at the forefront of modern science \cite{perra2015modeling,weng2013virality,Ronen30122014,quevillon2015social,petri2014homological}. Advances in technology continue to promote the accumulation of data, providing an optimistic light in the quest to understand these hugely complex systems. The task then, for researchers across a range of disciplines, is to find optimal ways to measure, model, analyze, and present the vast information at their disposal.

Network theory has proved to be an invaluable resource to exploit data on a large scale. Its great utility comes partly from the its ability to translate problems into a language independent of the particular subject of study. Hence, a ``node'' can represent entities as diverse as a human, a protein or a word \cite{granovetter1973strength,jeong2001lethality,motter2002topology}. ``Edges'' can represent any sort of interaction between the nodes, and concepts such as percolation, diffusion, paths and walks can all serve as models for various processes observed in the real world.

It is remarkable whenever the methods developed for the analysis of one subject matter are applied to seemingly unrelated problems. This occurs frequently when networks are involved. For example, the preferential attachment model can explain the distribution of citations in scientific literature as well as the distribution of popularity in a social network \cite{de1976general,barabasi2002evolution}, the PageRank algorithm was developed to rank websites but can also measure the risk of cancer in humans \cite{gleich}. These universalities motivate us to search for ways to measure networks and classify them by their properties; if we have a good description of the network, then we have potentially described a part of the ``real world'' which we would like to understand, moreover, we also have the entirety of past research and all the accompanying tools developed to help attack the problem.
\subsection{Motivation for ``dynamic communicability''}
Transmissible disease is possibly the best example to demonstrate the versatility of network analysis. Ultimately the theoretical considerations of network epidemiology involve nodes, edges and some knowledge of the disease itself such as the transmission probability, recovery rate and so on. Transmission could occur from one person to another, from one location to another (e.g. connected by air travel), or between species, but in each case the models employed remain well within the confines of the network framework \cite{stehle2011simulation,colizza2006role,johnson2015infectious}. This also extends to computer viruses \cite{pastor2001epidemic}, Twitter hashtags and internet memes \cite{10.3389/fphy.2015.00079,wang2015concurrent}, and possibly even cultural transmission on an archeological time-scale \cite{knappett2008modelling}. Clearly there is much to be gained from having a grounded understanding of how things spread through a network regardless of what that particular network represents.

\begin{figure*}[t]
    \subfloat[The left panel shows the network at each time-step (above) and its corresponding adjacency matrix (below). On the right the same information is represented as a list of temporal edges. Shown also are the different possible ways to randomize (or shuffle) the columns. Notice that simultaneously shuffling any two columns yields the completely shuffled edge-list shown in (iv).\label{big_fig}]{\includegraphics[width=\textwidth]{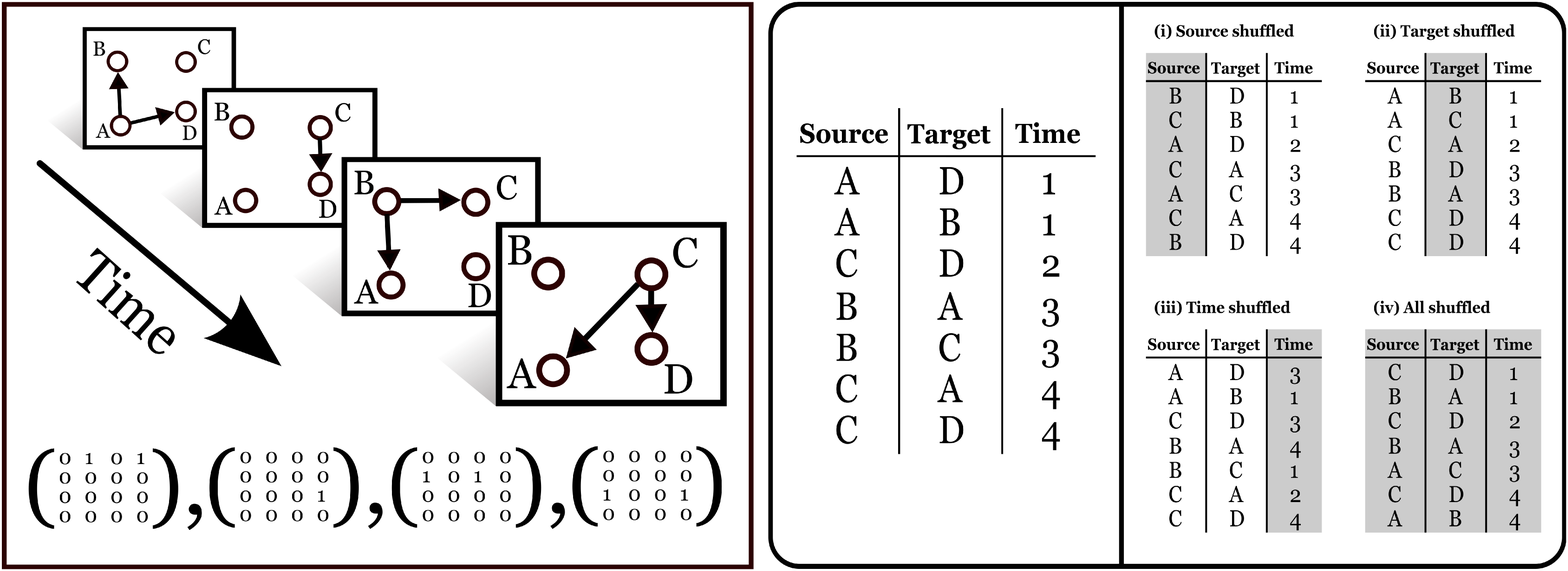}}\\

\subfloat[Each marker corresponds to a node in the example network. Each node is given a rank according to its broadcast score (left) and receive score (right). These rankings are plotted against the outgoing and incoming degree ranks respectively. The diagonal line divides the nodes into those that acheive higher broadcast (or receive) scores than expected, and those that are lower. \label{degree_rank_example}]{
        \includegraphics[width=0.23\textwidth]{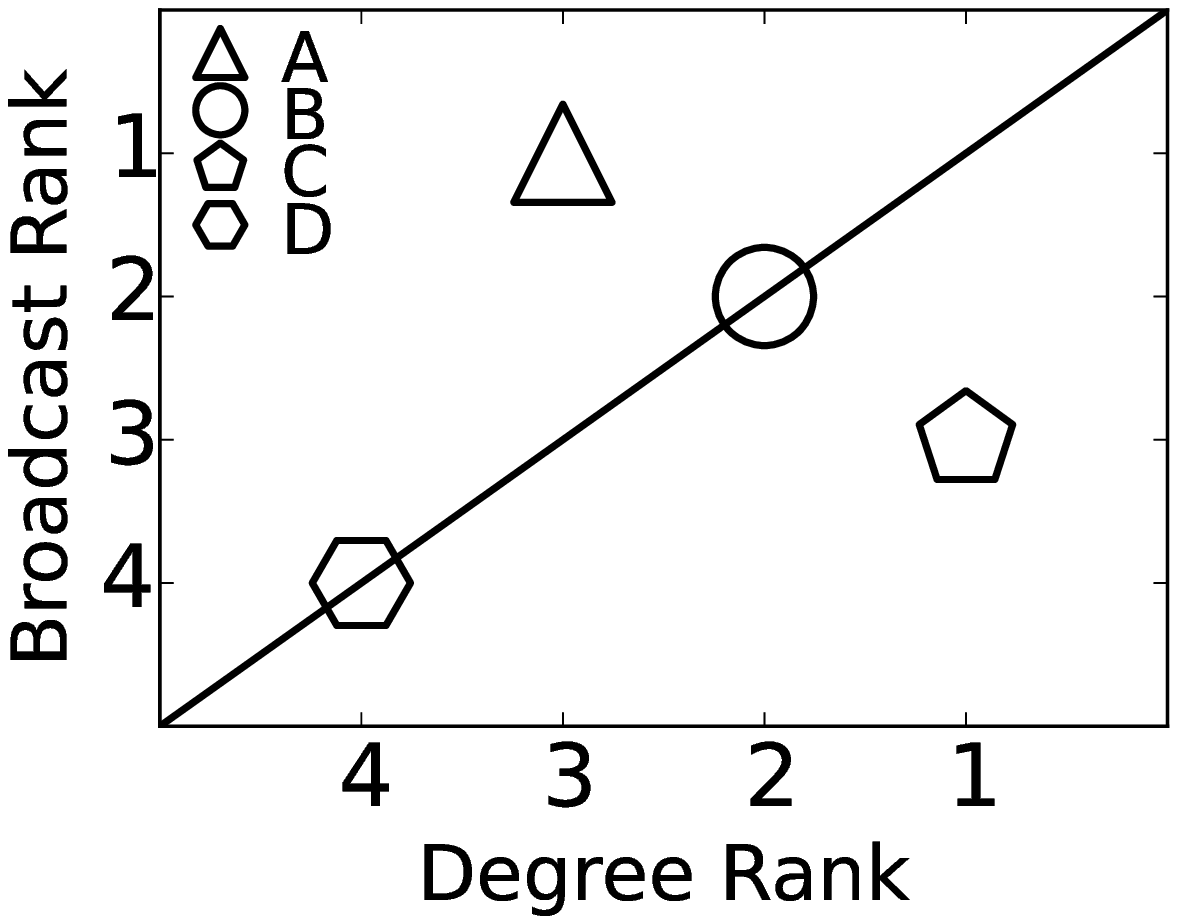}
        \includegraphics[width=0.23\textwidth]{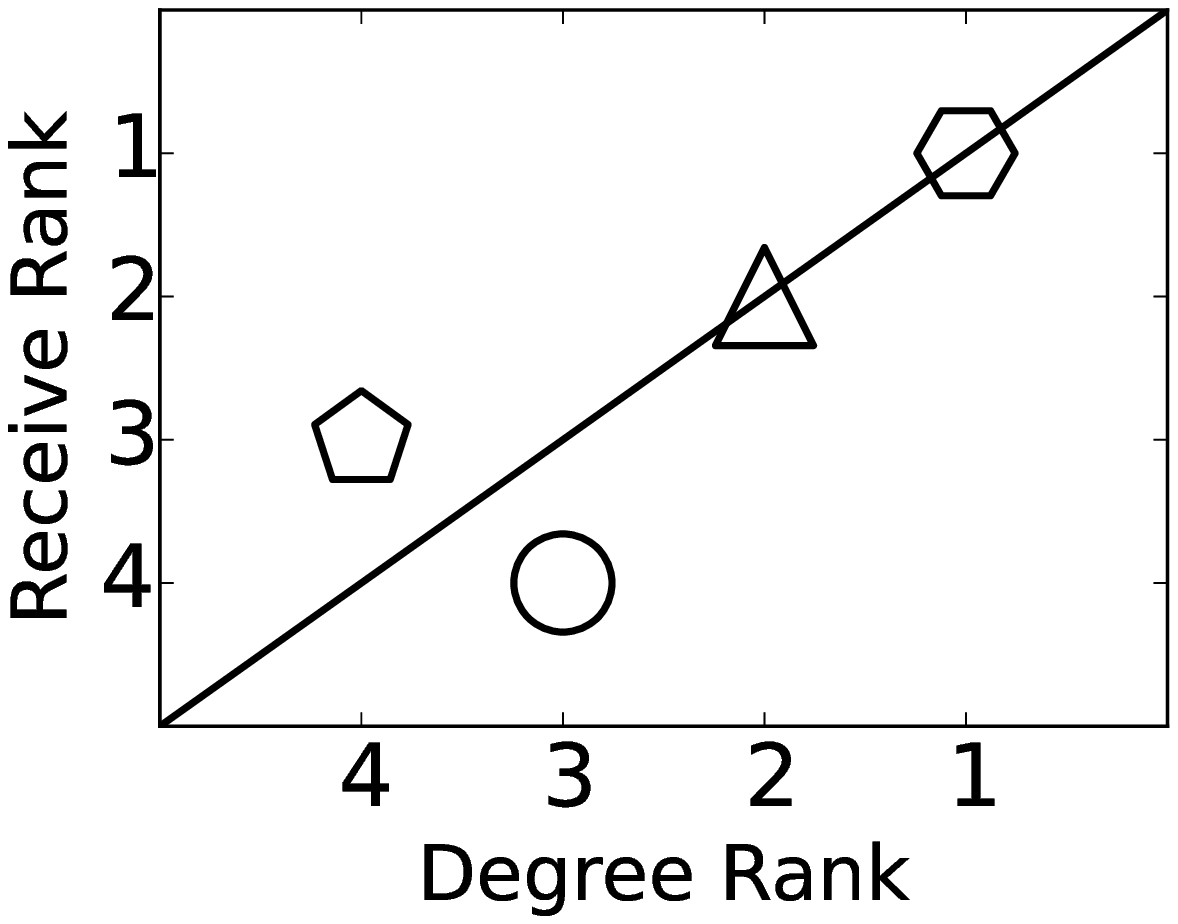}
				}\qquad
				\subfloat[The expectation values for each shuffling are calculated and the corresponding rank is plotted (we have chosen only to consider the target shuffling for broadcast score and source shuffling for receive score). The actual scores are shown by the darkness of the markers. \label{databases}]{
        \includegraphics[width=0.23\textwidth]{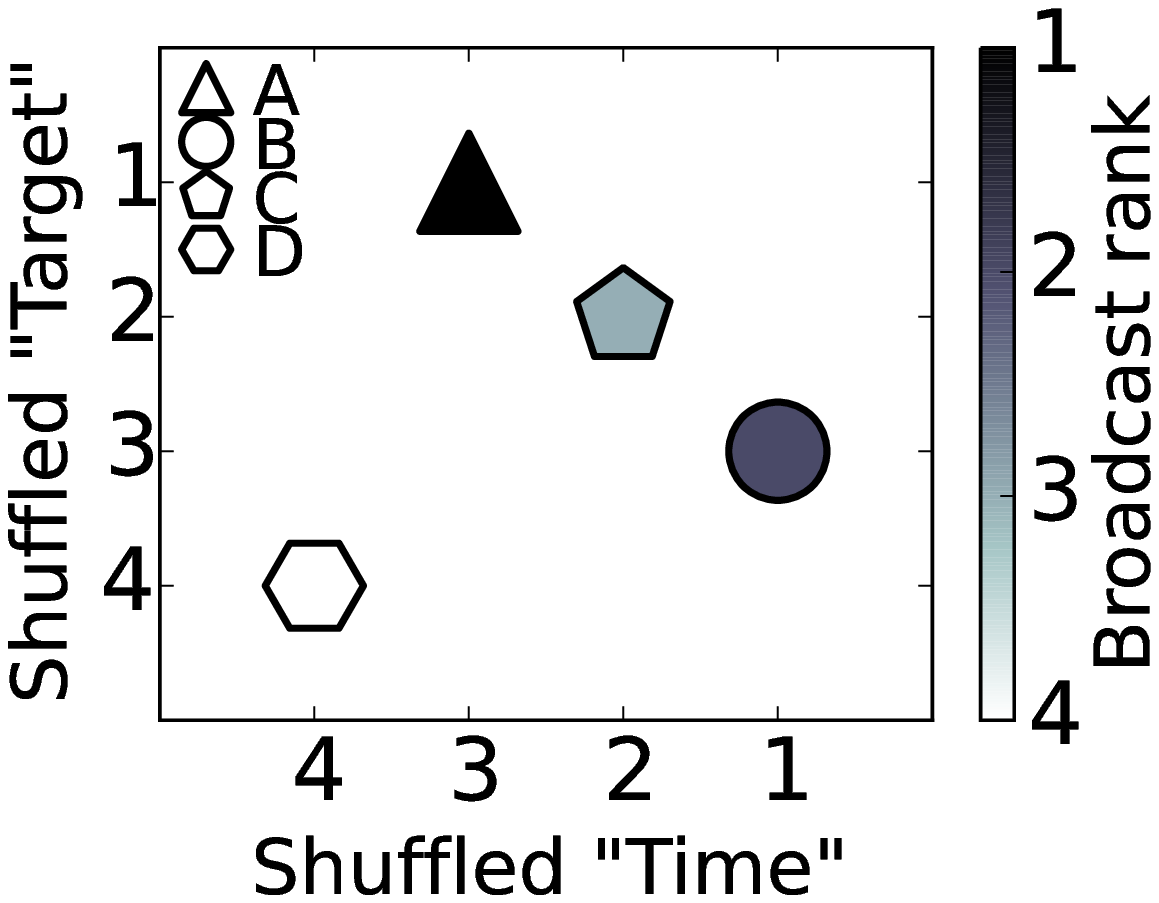}
        \includegraphics[width=0.23\textwidth]{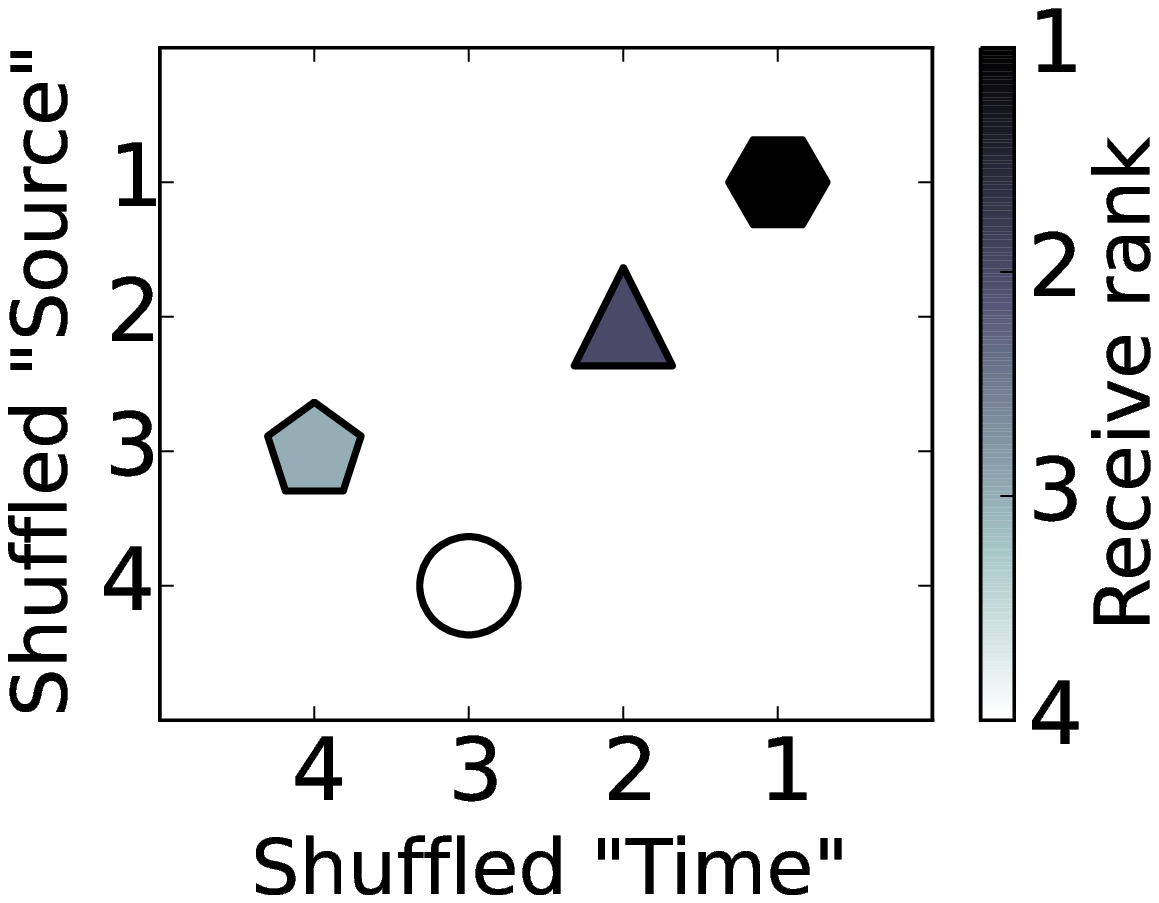}
				}
				\caption{A simple example of a directed temporal network. This example has been designed to illustrate the core concepts of this work. In Fig.(\ref{degree_rank_example}) it is apparent that the dynamic communicability of a node is not necessarily determined by its overall activity. Fig.(\ref{databases}) demonstrates how the dynamic communicability metrics can be broken down into temporal and structural elements. We apply the same visualization method to two real-world data-sets in Figs. (\ref{plots_2}) and (\ref{plots}).}
\end{figure*}

The work we present here concerns a scenario where we are given a database containing a set of distinct individuals, a set of pairwise interactions, and the exact time at which each interaction happened (see Fig.\ref{big_fig}). Additionally it is assumed that some transmissible agent was, or potentially could have been, spreading through the network. A practical question which often arises is: ``which node is potentially the most significant when it comes to the spread of a transmissible agent?''. 

To find the most influential spreader, given data of past interactions, there are several options to consider: the simplest method would be to find the individual with the highest node degree (this could be defined as either number of interactions that person had, or the number of people with whom they interacted). Alternatively we could use global network properties such as the betweenness centrality or closeness centrality of a node, both of which are defined on temporal networks \cite{holme2015modern}. The most extensive approach currently being used is to build a computational model of the process, adding as many factors into the model as one sees fit; where uncertainty is present, random variables can be used; and the centrality of an individual can be computed by running the model repeatedly and counting the proportion of simulations in which they are infected \cite{stehle2011simulation,richardson2015beyond}. 

Dynamic communicability, which was introduced in \cite{grindrod2011communicability} and is described in detail here in Section \ref{score_description}, offers a balance between the approach of modeling an epidemic-like process on a network, and simply measuring the size and shape of a network. Here we determine the influence of a node by counting the number of time-respecting walks that began at the node in question. In essence, we are using a model which assumes that a transmissible agent moves from one node to another at the exact time that an interaction takes place (which is known from the data) and with a given transmission probability. The fact that it is a walk (as opposed to a path) means that the agent can revisit previously infected nodes. Assuming this, and supposing that the pathogen is administered at node $i$, the broadcast score of $i$ tells us how large the expected outbreak will be. Supposing the pathogen is administered to a random unknown node, the receive score of $i$ tells us how likely that pathogen is to reach $i$.

\subsection{Separating dependencies}
In this paper we interrogate the two dynamic communicability metrics: broadcast score and its opposite, receive score. Through theoretical approaches we will examine how these centrality measures respond to different temporal network structures. Further, we derive methods to deconstruct the dynamic communicability measures into ``time dependent'' and ``structure dependent'' components. The formulae we derive achieve the same result as ``shuffling'' (randomly permuting) either the structural or temporal columns of the temporal edge-list respectively. This is an increasingly common technique used to determine the importance of various relationships within a database \cite{masucci2009differences,sanli2015local,karsai2011small,richardson2015beyond}. Here we employ this technique to unpick, from the information available, the factors most relevant to determining the outcome of a contagion-like process.

The following section explains in detail the dynamic communicability metrics. In Section \ref{model} we describe a stochastic model which can be tuned to reproduce various properties of the data. The main results from the model are a set of ``shortcut formulae'' for decomposing the dynamic communicability metrics into time dependent and structure dependent elements in an efficient way. We demonstrate these results on two publicly available data sets, which are described in Section \ref{data}, and the results are presented in Section \ref{results}. Section \ref{discussion} summarizes the findings from this work which we consider most significant.

\section{Definitions of ``Broadcast score'' and ``Receive score''}
\label{score_description}
Communicability, as introduced in \cite{estrada2008communicability}, is a measure of centrality based on the concept of ``walking'' on a network. A walk is any sequence of nodes in which one entry may only follow another if there is an edge in the network which connects them (if the network is directed then consecutive entries must follow the direction of the edge). The extension to temporal networks, in  which edges exist only at specified temporal instances, was introduced in \cite{grindrod2011communicability} and further developed in \cite{estrada2013communicability} and \cite{grindrod2014dynamical}. When dealing with temporal edges, we consider node sequences in which consecutive nodes are connected by an edge and, additionally, the time of that edge is later than (or at the same time as) its predecessor. These are referred to as ``time-respecting'' walks. 

Based on this premise it is possible to quantify the relationship between any two nodes: the ``dynamic communicability'' from node $i$ to node $j$, denoted $Q_{i,j}$, is a measure of the relative likelihood that a random walker injected into the network at $i$ will eventually pass through $j$. If we let $\theta_{i,j}^{(k)}$ be the number of time-respecting walks of length $k$ that begin at $i$ and end at $j$, then
\begin{equation}
\label{grindy_style}
Q_{i,j}=\sum_{k=0}^{\infty}\alpha^{k}\theta_{i,j}^{(k)}.
\end{equation}
The value $\alpha$ here is analogous to the probability of transmission (across an edge) in an epidemic spreading process. When chosen to be sufficiently small, it ensures that long walks are discounted heavily while short walks contribute more to the dynamic communicability metric. Several alternative approaches have been proposed as ways to down-weight walks based on their length. The original communicability defined on a static network discounts walks of length $k$ by dividing their number by $k!$, whereas when the temporal equivalent was introduced the exponential discounting shown in Eq.\eqref{grindy_style} was used. The measure introduced in \cite{estrada2013communicability} combines both. The measure introduced in \cite{grindrod2014dynamical} extends Eq.\eqref{grindy_style} by additionally discounting walks according to their duration in time.

For the first centrality measure, known as the ``Broadcast score'' of a node $i$, we compute the sum of all the discounted walks that begin at $i$ ($b_{i}=\sum_{j \in N} Q_{i,j}$). Similarly, to compute the second centrality measure, known as the ``Receive score'' of a node $i$, we sum all of the discounted walks that end at $i$ ($c_{i}=\sum_{j \in N} Q_{j,i}$).
  
\section{The model}
\label{model}
We use a simple yet versatile stochastic model to generate temporal networks. The parameters of the model can be manipulated to create synthetic data with properties similar to a wide range of temporal networks including those observed in many real interactive systems. Let there be $N$ nodes. The model proceeds over a series of discreet time-steps $\tau\in\{t_{0},t_{0}+1,...,t_{\text{end}}\}$ by the following rule:
\begin{quotation}
\noindent
At time $\tau$, with probability $\rho_{i,j}(\tau)$, a directed edge exists from node $i$ to node $j$. 
\end{quotation}

The adjacency matrix at time $\tau$,  $A_{\tau}$, will have a $1$ in location $i,j$ with probability $\rho_{i,j}(\tau)$ and be $0$ otherwise (it might often be the case that $A_{\tau}$ will be a matrix of zeros). The dynamic communicability matrix, as introduced in \cite{grindrod2011communicability}, over the sample (starting at $t_{0}$ and ending at $t_{\text{end}}$) is given in general by 
\begin{equation}
\label{com_def}
Q=(I-\alpha A_{t_{0}})^{-1}(I-\alpha A_{t_{0}+1})^{-1}\ldots(I-\alpha A_{t_{\text{end}}})^{-1}
\end{equation}
where $I$ denotes the identity matrix. But, as suggested in \cite{grindrod2013matrix}, we do not want to count paths that take multiple moves in a single time-step, so we will instead look at the variant definition
\begin{equation}
\label{Q_prod}
Q=(I+\alpha A_{t_{0}})(I+\alpha A_{t_{0}+1})...(I+\alpha A_{t_{\text{end}}}).
\end{equation}
Eqns. (\ref{com_def}) and (\ref{Q_prod}) are equivalent when $A_{\tau}^{2}=0$ for all $\tau$ (as this is the only way $(I-\alpha A_{\tau})^{-1}=(I+\alpha A_{\tau})$ can be true) i.e. provided that no walks of length 2 ever exist within a single time slice. Under these conditions our analysis also applies to the version of dynamic communicability defined in \cite{estrada2013communicability}, where $Q=\exp(\alpha A_{t_{0}})\exp(\alpha A_{t_{0}+1})...\exp(\alpha A_{t_{\text{end}}})$, since $\exp(\alpha A)=(I+\alpha A)$ when $A^{2}=0$.

The  time-dependent matrix 
\begin{equation}
P(\tau) =
 \begin{pmatrix}
  \rho_{1,1}(\tau) & \rho_{1,2}(\tau) & \cdots & \rho_{1,N}(\tau) \\
  \rho_{2,1}(\tau) & \rho_{2,2}(\tau) & \cdots & \rho_{2,N}(\tau) \\
  \vdots  & \vdots  & \ddots & \vdots  \\
  \rho_{N,1}(\tau) & \rho_{N,2}(\tau) & \cdots & \rho_{N,N}(\tau)
 \end{pmatrix},
\end{equation}
to a large extent, describes the entire structure of the network and its evolution over time. Our approach to exploring dynamic communicability of networks generated by this model involves considering the various forms that $P$ can take; then examining the expectation of $Q$ as we iteratively increase the number of terms on the right hand side of Eq.\eqref{Q_prod}. The following analysis requires that the values contained in $P$ are small enough that the probability of generating a matrix containing a path of length $2$ or more is negligible. The model is therefore more applicable to temporally highly resolved data-sets as opposed to those in which a relatively small number of temporal instances are recorded. 

\subsection{Receive score}
If we think about constructing $Q$ iteratively i.e. starting at time $t_{0}$ with $Q_{0}=(I+\alpha A_{t_{0}})$, then multiplying on the right by $(I+\alpha A_{t_{0}+1})$, then again by the next term, then the next etc., then
\begin{equation}
\label{iter}
Q_{t}=Q_{t-1}\times(I+\alpha A_{t_{0}+t}),
\end{equation}
where $t$ indexes the number of times the iteration has been performed. After $t=t_{\text{end}}-t_{0}$ iterations we have the desired $Q_{t}=Q$. The effect of one iteration can be seen on a $3\times 3$ example:
\begin{equation}
\scriptsize
 \begin{pmatrix}
  q_{1,1} & q_{1,2}  & q_{1,3} \\
  q_{2,1} & q_{2,2}  & q_{2,3} \\
  q_{3,1} & q_{3,2}  & q_{3,3}
 \end{pmatrix}
  \begin{pmatrix}
  1 & 0  & 0 \\
  0 & 1  & \alpha \\
  0 & 0  & 1
 \end{pmatrix}
 = \begin{pmatrix}
  q_{1,1} & q_{1,2}  & q_{1,3}+\alpha q_{1,2} \\
  q_{2,1} & q_{2,2}  & q_{2,3}+\alpha q_{2,2} \\
  q_{3,1} & q_{3,2}  & q_{3,3}+\alpha q_{3,2}
 \end{pmatrix}.
\end{equation}
\normalsize
In general, provided $A_{\tau}^{2}=0$ for all $t_{0}<\tau\leq t_{\text{end}}$, if the $i,j$th entry of $A_{\tau}$ is $1$ then the $i$th column is multiplied by $\alpha$ and added to the $j$th column. Since the receive score after $t$ iterations is equal to the (row) vector of column sums of $Q_{t}$,
\begin{equation}
\mathbf{c}(t)=(c_{1}(t),c_{2}(t),...,c_{N}(t)),
\end{equation} 
we can describe its evolution as $t$ increases as follows: at each iteration choose $i$ and $j$ with probability $\rho_{i,j}(t_{0}+t)$ and update by setting
\begin{equation}
\label{algo}
\mathbf{c}(t+1)=(c_{1}(t),c_{2}(t),...,c_{j}(t)+\alpha c_{i}(t),...,c_{N}(t)).
\end{equation}
In matrix notation this is 
\begin{equation}
\label{c_evo}
\mathbf{c}(t+1)=\mathbf{c}(t)(I+\alpha A_{t_{0}+t}).
\end{equation}
\subsection{Expectation of receive score}
\label{expectation}
The receive score is dependent on $t$. To examine this dependence, we focus on the expectation of $c_{i}(t)$, denoted $\hat{c}_{i}(t)$, which is computed by taking the mean over many networks generated by the described model for some given $P$. For analytical considerations we assume that all of the $c_{i}(t)$ are well approximated by their mean. A similar approach is found in \cite{rogers2014null}. The growth of $\hat{c}_{i}(t)$ is then described by
\begin{equation}
\hat{c}_{i}(t+1)=\hat{c}_{i}(t)+\alpha \sum_{j=1}^{N}\rho_{j,i}(t_{0}+t)\hat{c}_{j}(t).
\end{equation}
The right hand side here equation sums over all possible changes that can happen to $c_{j}$ and their associated probabilities. This is equivalent to replacing $A_{t_{0}+t}$ in Eq.(\ref{c_evo}) with the expectation of $A_{t_{0}+t}$, which happens to be $P(t_{0}+t)$. We have 
\begin{equation}
\mathbf{\hat{c}}(t+1)=\mathbf{\hat{c}}(t)[I+\alpha P(t_{0}+t)].
\end{equation} 
For large time-scales, we can say that $\hat{c}_{j}(t+1)-\hat{c}_{j}(t)\approx \partial \hat{c}_{j}/\partial t$, giving
\begin{equation}
\label{diff_c}
\frac{\partial \mathbf{\hat{c}}(t)}{\partial t}=\alpha\mathbf{\hat{c}}(t)P(t_{0}+t).
\end{equation}
An almost identical derivation can be performed to find a similar expression for $\mathbf{\hat{b}}$. In this case, instead of starting the iterative process at $t_{0}$ and multiplying on the right, as in Eq.\eqref{iter}, we start at time $t_{\text{end}}$ with $Q_{0}=(I+\alpha A_{t_{\text{end}}})$ and iterate by multiplying on the left, i.e $Q_{t}=(I+\alpha A_{t_{\text{end}}-t})\times Q_{t-1}$. Following similar steps we arrive at
\begin{equation}
\label{diff_b}
\frac{\partial \mathbf{\hat{b}}(t)}{\partial t}=\alpha P(t_{\text{end}}-t)\mathbf{\hat{b}}(t)
\end{equation}
where $\mathbf{\hat{b}}(t)$ is a column vector of the expectation of the broadcast scores. Our theoretical results stem from these two equations, solutions can be found for various forms of $P(\tau)$, here we mention a few simple cases.
\subsection{Time-independent $P$ matrix}
When $P$ is a constant matrix, the (well known) general solution to Eq.(\ref{diff_c}) is
\begin{equation}
\label{c_sol}
\mathbf{\hat{c}}(t)=\mathbf{\hat{c}}(t_{0})e^{\alpha Pt}
\end{equation}
where 
\begin{equation}
\label{powers}
e^{\alpha Pt}=\sum_{k=0}^{\infty}\frac{1}{k!}(\alpha Pt)^{k}.
\end{equation}
\subsubsection{Equivalence to shuffling the time column}
Consider a temporal edge-list where the ``time'' column has been shuffled as shown in part (iii) of Fig.(\ref{big_fig}). The overall number of interaction events between each pair of nodes is unchanged, however each of these events now occurs at some random point in time. The time-series of interaction events from node $i$ to node $j$ can be modeled by a Bernoulli process, i.e. at each discreet time-step there is a fixed probability that an edge from $i$ to $j$ will exist. If we have a sufficiently large amount of data then the matrix of these time-independent probabilities, which happens to be $P$, can be approximated easily as we show in this section. The above result can then be used to predict the dynamic communicability metrics of the time-shuffled edge-list.

We can infer $P$ from the data by constructing a weighted adjacency matrix $W$ where $W_{i,j}$ is the total number of times each edge appears in the temporal edge-list. To infer a time-independent probability $\rho_{i,j}$ that an edge exists at time $\tau$ (for any $t_{0}\leq \tau < t_{\text{end}}$) we normalize by the number of time steps in the sample:
\begin{equation}
\label{normed}
 \rho_{i,j}=\frac{W_{i,j}}{t_{\text{end}}-t_{0}}.
\end{equation}
Since $t_{0}-1$ lies outside the time for which data is sampled, $A_{t_{0}-1}$ is a zero matrix, giving $\mathbf{\hat{c}}(t_{0}-1)=\mathbf{1}$  where $\mathbf{1}$ is a row vector of length $N$ and all entries are $1$. Substituting the $P$ matrix associated with Eq.\eqref{normed} into Eq.\eqref{c_sol}, and into the equivalent result for $\mathbf{\hat{b}}$, we arrive at the concise formulae for computing the expectation of the broadcast and receive scores of a time-shuffled network,
\begin{equation}
\label{W_sol_b}
\mathbf{\hat{b}}=e^{\alpha W}\mathbf{1}
\end{equation}
and
\begin{equation}
\label{W_sol_c}
\mathbf{\hat{c}}=\mathbf{1}^{T} e^{\alpha W}
\end{equation}
respectively. A very fast open-source algorithm for solving the matrix exponential for large matrices has recently been developed \cite{al2009new}. Applying this method gives a prediction for the outcome of averaging a large number of shuffled temporal edge-lists where the ``time'' column has been shuffled. The comparison between the prediction and the actual shuffled data is shown in Fig.(\ref{shuffle_test}).
\subsubsection{Heterogenous ``send'' and ``receive'' model}
\label{sr_section}
Consider a temporal edge list for which all three columns have been shuffled as in part (iv) of Fig.(\ref{big_fig}). While much of the relational information will be lost, the number of times each node is found in the ``source'' column will be unchanged and therefore the outgoing degree of each node is retained, similarly the incoming degree is unchanged by the shuffling of the ``Target'' column. This process bears much resemblance to the configuration model of \cite{molloy1995critical} in which each node has a given degree but the pairwise connections are randomized. Related models, which replace the exact degree sequence with a sequence of fitness variables (giving the propensity of each node to attract edges), have been studied \cite{chung2002connected}; this happens to be a case where Eqs.\eqref{W_sol_b} and \eqref{W_sol_c} can be solved analytically. 

Let $s_{i}$ be the probability that node $i$ has an outgoing edge in any given time-step (we have chosen the letter $s$ as this represents the `sending' of information), and let $r_{i}$ be the probability that $i$ has an incoming edge in any given time-step ($r$ to represent the `receiving' of information). With the vector notation, $\mathbf{s}=[s_{1},s_{2},\ldots,s_{N}]^{T}$ and $\mathbf{r}=[r_{1},r_{2},\ldots,r_{N}]^{T}$, we have 
\begin{equation}
P=\mathbf{s}\mathbf{r}^{T}.
\end{equation}
We add the condition that $\sum_{i} s_{i}=\sum_{i} r_{i}=1$ then the expected number of edges per time-step is $1$ (meaning that when comparing to data we can treat $t$ as the total number of interactions). Under these conditions the solution to Eq.\eqref{diff_c} (see Appendix \ref{sr_sol}) is
\begin{equation}
\label{c_time_dep}
\mathbf{\hat{c}}(t)=\mathbf{1}+\frac{e^{\alpha [\mathbf{r}^{T}\mathbf{s}]t}-1}{\mathbf{r}^{T}\mathbf{s}}\mathbf{r}^{T}.
\end{equation}
Two main conclusions come from this result: firstly, the receive score of a node is proportional to its propensity to attract incoming edges (for broadcast score it is the outgoing edges, see Appendix \ref{sr_sol}). Second, as the sample size $t$ increases the score increases exponentially.
\subsection{Time-dependent $P$ matrix}
A general solution to Eq.\eqref{diff_c} for any $P(\tau)$ does not exist, we instead incorporate a limited amount of temporal information by expanding the ``send'' and ``receive'' model of the previous section. Suppose we have the model from Section \ref{sr_section} with the modification that the ``receive'' vector $\mathbf{r}$ is now a function of time, say $\mathbf{r}(\tau)$, then Eq.\eqref{diff_c} reduces to
\begin{equation}
\label{time_s}
\frac{\partial \mathbf{\hat{c}}(t)}{\partial t}=\alpha \mathbf{r}(t_{0}+t)e^{\alpha \mathbf{s}^{T}\int_{0}^{t}\mathbf{r}(t_{0}+t') d t'}
\end{equation}
(see Appendix \ref{t_model}). Eq.\eqref{time_s} allows us to examine special cases where the order in which messages are sent affects the receive score of each node.
\subsubsection{Simple time-dependent example}
\label{simple_time}
Before we derive a result applicable to real-world data, we introduce a simple example to provide some intuition for the time-dependence. We consider the case where each node is active only once during the duration of the sample. Suppose node $i$ receives $r_{i}$ edges at time $\tau_{i}$. We can write the corresponding $r$ vector using the Dirac $\delta$:
\begin{equation}
r_{i}(\tau)=r_{i}\delta(\tau-\tau_{i}).
\end{equation}
The justification for this choice of $\mathbf{r}(\tau)$ is that the expected number of messages received by $i$ over some interval will be $r_{i}$ if the time interval includes $\tau_{i}$. For convenience we suppose, without loss of generality, that $\tau_{i}=i$ for all $i\in 1,2,...,N$. In Appendix \ref{simple_time2} Eq.\eqref{c_time_dep} is solved with this form of $\mathbf{r}(\tau)$ to get
\begin{equation}
\hat{c}_{i}(N)=1+\alpha r_{i}\exp\left(\alpha\sum_{j=0}^{i}r_{j}s_{j}\right).
\end{equation}
This result shows that nodes which interact later in the sample will have, on average, exponentially higher receive scores. In a similar way, it can be shown that a node which acts earlier in the sample has an exponentially higher broadcast score.

\subsubsection{Incorporating empirical data}
\begin{figure}[H]
		\includegraphics[width=0.47\textwidth]{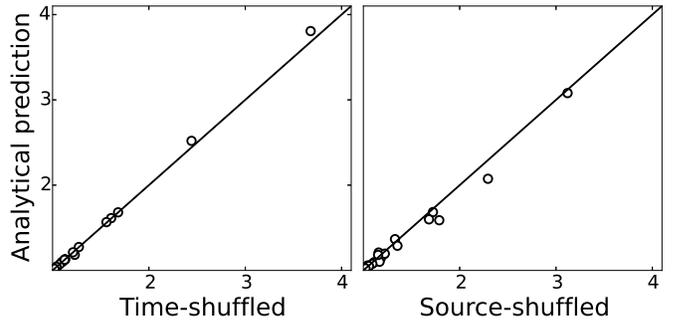}
    \caption{A demonstration of the accuracy of the derived formulae using a sample of $23$ nodes from the Enron data-set (all of which have at least one outgoing edge within the sample) and a total of $312$ emails. Each marker represents an employee. In both plots the $x$-axis shows the receive score computed by shuffling one column of the edge-list, as shown in Fig.(\ref{big_fig}), and averaged over $100$ shuffles. In the left hand plot the time column was subjected to shuffling and $y$-axis shows the receive score as predicted by Eq.(\ref{W_sol_c}), in the right the target column was subjected to shuffling and $y$-axis shows the receive score as predicted by Eq.(\ref{time_formula}). $\alpha=0.02$.}
		\label{shuffle_test}
\end{figure}

Suppose that for each node $i$ we know the time of every received edge but do not know where the edge originated from (this corresponds to the source shuffled network). We can achieve this by choosing
\begin{equation}
\label{r_delta}
r_{i}(\tau)=\sum_{k\in K_{i}}\delta(\tau-\tau_{i}^{(k)})
\end{equation}
where $K_{i}$ is the set of edges for which $i$ is the target and $\tau_{i}^{(k)}$ is the time at which edge $k$ was present. More important, however, is the function $R_{i}(\tau)$ which we define as the number of messages that have been received by $i$ between $t_{0}$ and $\tau$, and can be expressed as
\begin{equation}
R_{i}(\tau)=\int_{t_{0}}^{\tau}r_{i}(t')dt'. 
\end{equation}
To achieve the correct normalization (for the expectation of the total number of edges to agree with the data) we choose $s_{i}$ to be the probability that any given edge is sent from $i$, this is inferred using  
\begin{equation}
s_{i}=\frac{\sum_{j}W_{i,j}}{\sum_{i,j}W_{i,j}}.
\end{equation}
The solution to Eq.\eqref{diff_c}, which we derive in Appendix \ref{source_shuffled_derivation}, is 
\begin{equation}
\label{time_formula}
\hat{c}_{i}=1+\alpha\sum_{k\in K_{i}}\exp\left(\alpha\sum_{j\in N}s_{j}R_{j}(\tau_{i}^{(k)})\right).
\end{equation}
This formula predicts the average of the receive score over many networks generated by shuffling the Source column in the original data. The analytical prediction and average shuffling results are shown in Fig.(\ref{shuffle_test}). In our data analysis we also use an equivalent formula to predict the outcome of shuffling the target column and calculating the broadcast score. The derivation is similar to that of Eq.\eqref{time_formula}. We get
\begin{equation}
\label{time_formula_b}
\hat{b}_{i}=1+\alpha\sum_{k\in \kappa_{i}}\exp\left(\alpha\sum_{j\in N}r_{j}S_{j}(\tau_{i}^{(k)})\right)
\end{equation}
where $S_{j}(\tau)$ is the number of messages that have been sent by $i$ between $\tau$ and $t_{\text{end}}$, $r_{j}$ is the time-independent probability that $j$ receives a message in any given time-step, and $\kappa_{i}$ is the set of edges for which $i$ is the source.
\section{Data}

\label{data}
\subsection{Enron}
We downloaded the entire Enron email corpus that was made publicly available during an investigation by the Federal Energy Regulatory Commission into the events leading to its bankruptcy \cite{enron}. The data contains the mailing history of 150 Enron employees between 1999 and 2003. A folder exists for each of the named employees, each of which contain a number sub-folders, and each subfolder contains a number of text files; the text files contain the emails themselves and some meta-data. The naming of the folders is not consistent across employees; most sent emails belong to a folder labelled ``sent'', ``sent email'', or something similar but there are also many exceptions. A consistent format was found across all the text files with the time-stamp located on the first line, the ``From'' field appearing on the second, and the ``To'' field starting on the third line and often extending over several lines where emails have been sent to multiple recipients. 

We crawled every text file within sub-folders named ``sent'', ``sent\_items'' and ``\_sent\_mail', reading the specific lines which correspond to the ``From'' field, the ``To'' field and the time-stamp. Within the ``From'' and ``To'' lines we found all substrings which resemble a distinct email address i.e. bound on either side by blank spaces and contain the ``@'' symbol. From these data we constructed a temporal edge list of the form shown in Fig.(\ref{big_fig}) where the node IDs are email addresses. Multiple edges were created for emails with multiple recipients. In several cases the email addresses found in the ``From'' field, across the emails of an individual employee, would not always be identical. Usually this was because of the use of email aliases although on a small number of occasions  this was clearly not the case. At our own discretion, we replaced the node ID of all aliases relating to an employee with a single node ID.  

Many of the emails were sent to addresses outside of the corporation, these were removed from our data. We also found that some employees in the data-set had very little or no activity; we therefore reduced the sample to only those who have both sent, and received, at least one email to other users within the sample. After trimming, the network has $141$ nodes and a total of $21,303$ temporal edges.

We also incorporated information regarding the roles of each employee according to enron.org \cite{names}. The following abbreviations have been used for the legend in Fig.(\ref{enron_plot}): EMP$=$employee, TRA$=$trader, LAW$=$lawyer, MAN$=$manager, DIR$=$director, VP$=$vice president, MD$=$managing director, PRE$=$president, CEO$=$chief executive, ???$=$unknown.

The sample of emails we have chosen to use is by no means complete, however, it is our belief that the methods used to sample this data avoid introducing any biases which would compromise the results we present.
\begin{figure*}[t!]
\centering
    \subfloat[Enron email results \label{enron_raw}]{
		\includegraphics[height=6cm]{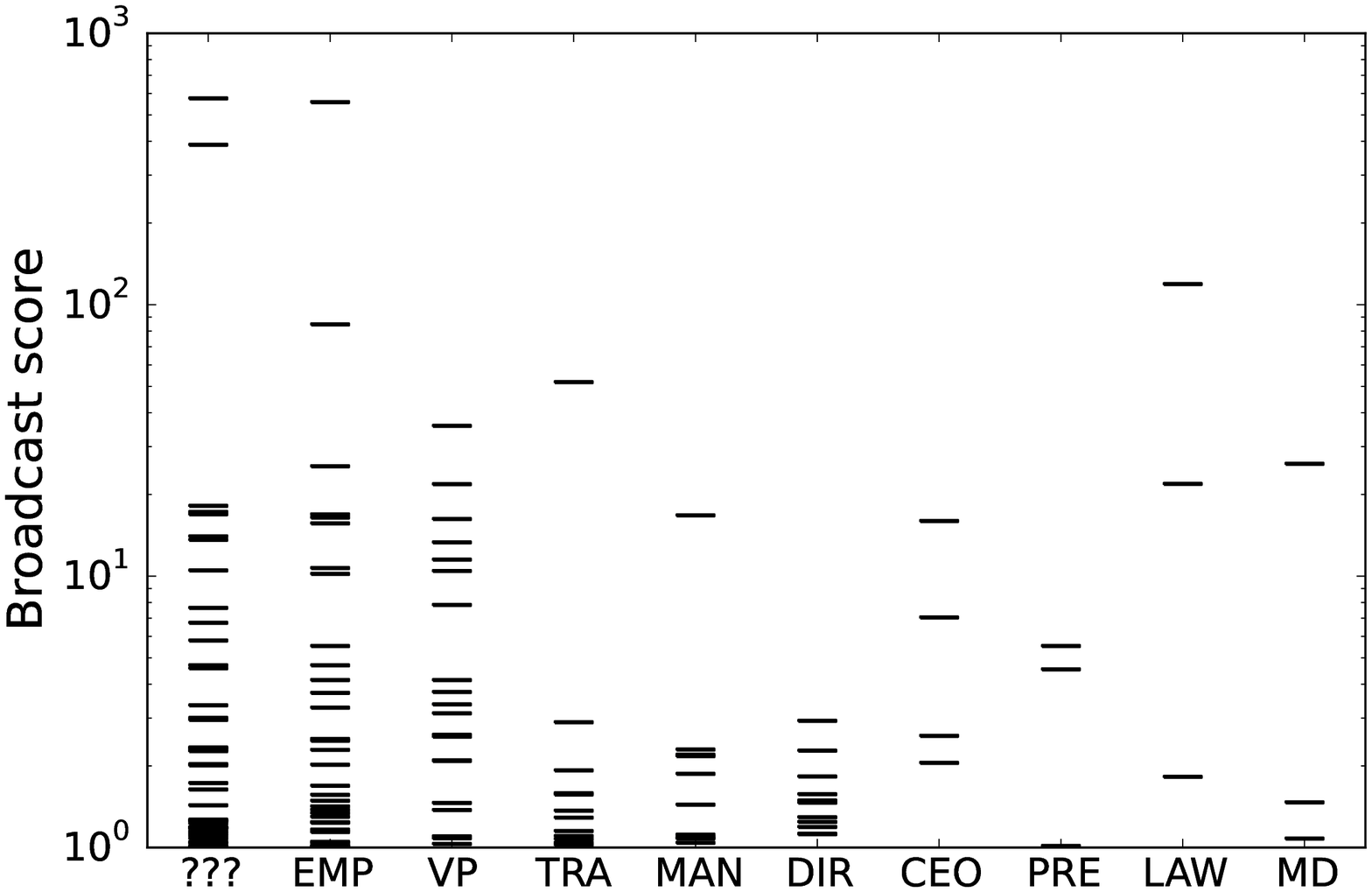}}\qquad
		    \subfloat[Hospital results \label{hospital_raw}]{
		\includegraphics[height=6cm]{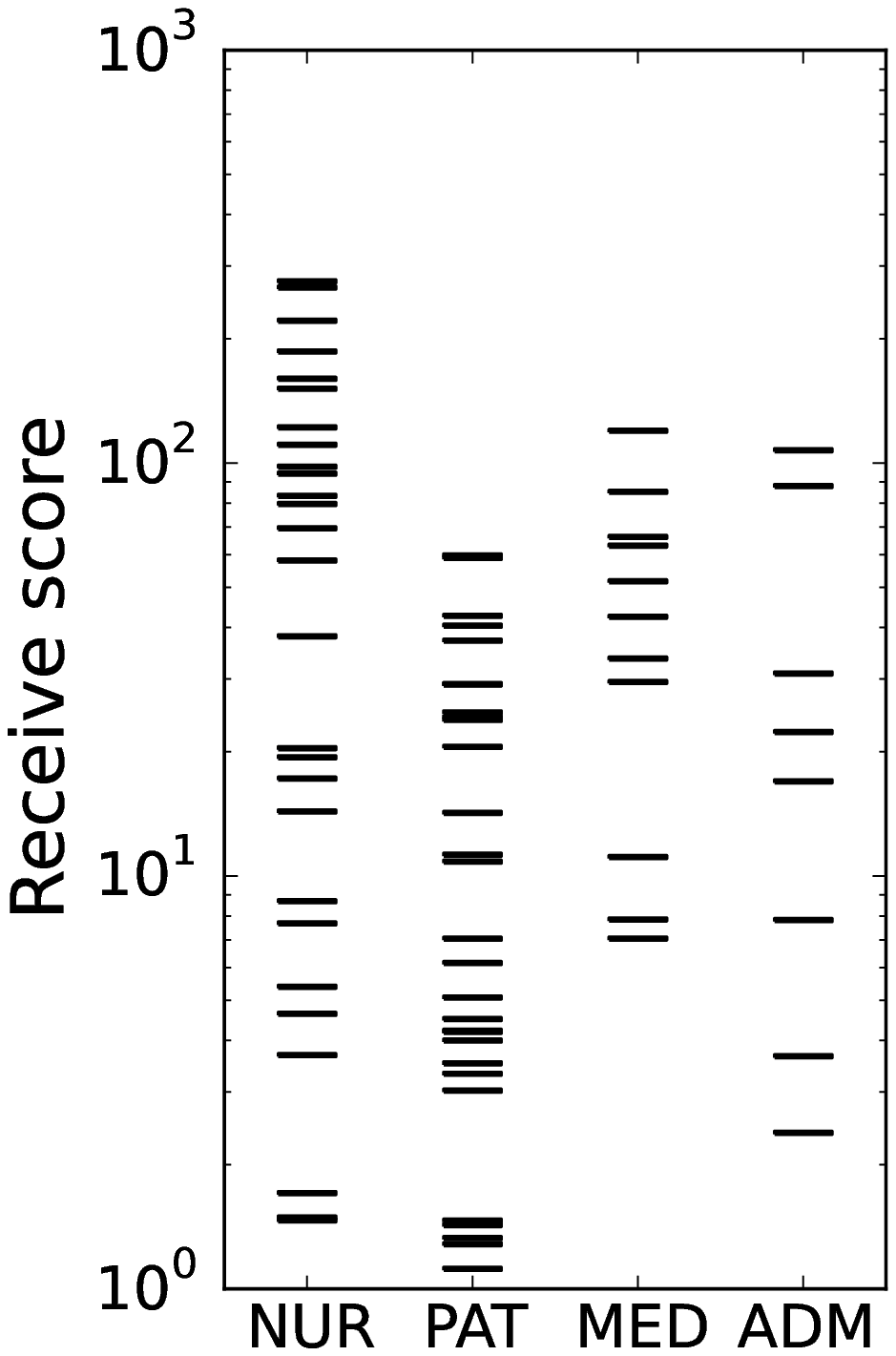}}
    \caption{Each marker (short horizontal line) represents an Enron employee or participant in the Sociopatterns experiment. The broadcast score, computed by Eq.\eqref{bc_compute}, is displayed on the $y$-axis in (\ref{enron_raw}) and similarly the receive score in (\ref{hospital_raw}) from Eq.\eqref{rc_compute} with $\alpha=0.01$ and $\alpha=0.005$ respectively. In both the employees (or participants) are divided into distinct categories shown along the $x$-axis. Abbreviations are given in Section \ref{data}.}
		\label{box}
\end{figure*}
\subsection{Sociopatterns hospital ward}
We downloaded the Hospital ward dynamic contact network from the Sociopatterns website (refer to \cite{10.1371/journal.pone.0073970} for details). The data was collected using proximity sensors attached to each participant. In the original data, every instance (instances are recorded every 20 seconds) in which two participants are ``interacting'' (i.e. within a given proximity of each other) is presented in a temporal edge list of the form shown in Fig.(\ref{big_fig}). Consequently, interactions which occur for a prolonged duration appear in the data multiple times so we performed the following reduction: where the same pair of participants were found to be interacting on multiple consecutive time-steps, all but one of the corresponding rows in the edge list were removed, leaving only the first of such instances. For each remaining row we create two edges in the processed temporal edge-list, one in each direction between the pair of participants interacting, both edges have the same time-stamp. Our analysis therefore considers transmission to occur at the first moment an interaction begins and does not depend on its duration. After processing, the network has $75$ nodes and a total of $28,076$ temporal edges.

\subsection{Algorithms}
\label{decision}
Much of the related literature formulates the problem of computing a dynamic communicability matrix using a series of linear algebra operations \cite{grindrod2011communicability}. This approach utilizes the adjacency matrix for the network at each time step (see Fig.(\ref{big_fig})) and assumes that within each time-slice the hypothetical random walker can traverse edges instantaneously, i.e. without requiring that time move forward for them to perform the movement. Consequently, if there is any cycle within a single time-slice (including for example an edge from $i$ to $j$ and another from $j$ to $i$) then there will be paths of infinite length, meaning that $\alpha$ must be restricted to a particular range of values to guarantee convergence \cite{greetham2013centrality}.

In this work we remove the assumption that a walk can traverse more than one edge per time slice (as suggested in \cite{grindrod2013matrix}). Moreover, we suggest the following recursive approach to computing the dynamic communicability metrics which avoids the need to perform any matrix operations. 

Suppose we have a network $G$ with each temporal edge denoted by a triple $(i,j,t)$ where $i$ is the source node, $j$ is the target node and $t$ is the time. Rewriting Eq.(\ref{c_evo}) with this notation we have  
\begin{equation}
c_{i}(\tau+1)=c_{i}(\tau)+\alpha\sum_{(i,j,\tau)\in G}c_{j}(\tau)
\end{equation}
with $c_{i}(t_{0}-1)=1$.
Then the receive score for node $i$ computed between time $t_{0}$ and $t_{\text{end}}$ is given by
\begin{equation}
\label{rc_compute}
c_{i}=c_{i}(t_{\text{end}}).
\end{equation}
Similarly, for the broadcast score we have
\begin{equation}
b_{i}(\tau-1)=b_{i}(\tau)+\alpha\sum_{(i,j,\tau)\in G}b_{j}(\tau)
\end{equation}
with $b_{i}(t_{end}+1)=1$. Then the broadcast score for node $i$ computed between time $t_{0}$ and $t_{\text{end}}$ is given by
\begin{equation}
\label{bc_compute}
b_{i}=b_{i}(t_{0}).
\end{equation}
If we were to first compute the vector $c_{i}(t_{0})$ for all the nodes $i$, then commit these values to memory, then compute $c_{i}(t_{\text{end}}+1)$ for all $i$, and continue in this fashion, then the addition operations we perform are precisely the same as those performed in the established matrix multiplication method \cite{grindrod2013matrix}. The advantage of this implementation, however, is that the score for a single node can be computed lazily, that is, without wasting unnecessary time. (It is important, when using this method, to use memoization to avoid repeating a large number of calls to the functions $c_{i}$ and $b_{i}$). Computationally we can be certain that these algorithms are at least as fast as the current alternatives.
\section{Results}
\label{results}
\subsection{Modeling}
In Section \ref{model} we derived formulae which predict the outcome of calculating the broadcast score for a large number of shuffled temporal edge-lists. The amount of error in these predictions is illustrated in Fig.(\ref{shuffle_test}) where we see that Eq.\eqref{W_sol_c} gives accurate results regarding temporal edge lists with the time-column shuffled. The corresponding result, Eq.\eqref{time_formula}, appears to be less reliable however, owing to the computational cost of calculating the receive score multiple times, we chose only to test a very small sample. This contradicts the assumptions of the analytical model; particularly the assumption made in Section \ref{expectation} that the score $c_{i}(t)$ in an individual generation of the probabilistic model is well approximated by its mean, at time $t$, over many generations. It is likely that in a small data-set that there is a high variance in the distribution of receive scores and we expect the prediction to improve as the number of interactions increases. The creation of these ``shortcut'' formulae allowed us to perform data analysis on two large scale temporal edge-lists which would have otherwise taken an inconvenient amount of computation.  
\begin{figure*}[t]
    \centering
    \subfloat[Enron email directory\label{enron_plot}]{
        \includegraphics[width=0.46\textwidth]{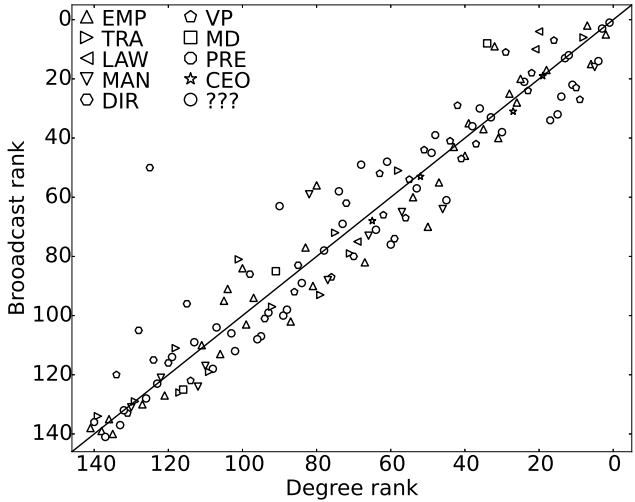}
				}	\quad
		\subfloat[Hospital contact network\label{hospital_plot}]{
        \includegraphics[width=0.46\textwidth]{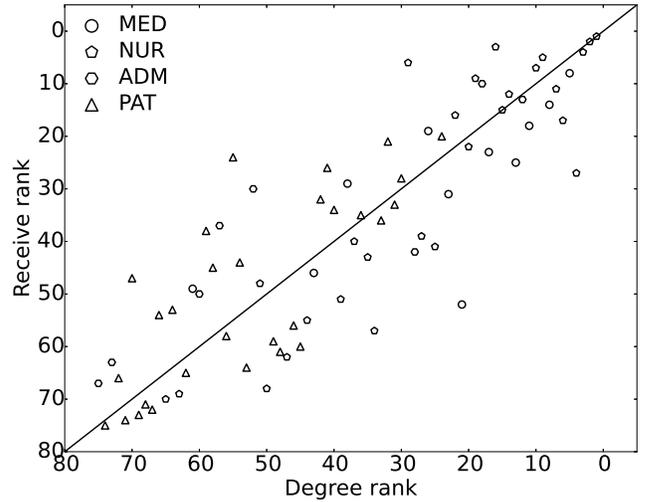}
	}
    \caption{The rank according to broadcast score (left, computed by Eq.\eqref{bc_compute}) and receive score (right, computed by Eq.\eqref{rc_compute}), with $\alpha=0.01$ and $\alpha=0.005$ respectively, plotted against the out-degree (left) and in-degree (right). Each individual in the network is represented by a data point, their classification is given by their shape. The abbreviations in the legend are explained in Section \ref{data}. The one-to-one line is plotted as a visual aid to partition the nodes into two groups; those which have higher than expected scores (top left), and those who have lower than expected (bottom right).}
\label{plots_2}
\end{figure*}
\subsection{Data analysis}
Using the method described in Section \ref{score_description} we calculated the broadcast score for the Enron email corpus and the receive score for the Sociopatterns hospital ward experiment. We have chosen values of $\alpha$ that produce visually interesting figures; when too small the calculation of broadcast and receive scores are dominated by the contribution from walks of length $1$ and therefore become equivalent to the out-degree and in-degree respectively. Conversely, when $\alpha$ is too large, long walks dominate the scores and the edges with early timestamps determine the outcome. To anyone considering using these methods we recommend that a range of $\alpha$ values be tested, each one potentially exposing different information about the behavior of the system. For an in-depth analysis of $\alpha$ and its interpretation see \cite{Chen21062016}. 

The results are presented first in Fig.(\ref{box}). In Fig.(\ref{plots_2}) we compare the result of each individual with their overall activity. We note two observations from Fig.(\ref{plots_2}): one Enron employee (a director) stands out as having an unusually high broadcast score when compared to a low amount of overall activity (broadcast rank $50$, degree rank $125$), and that patients in the hospital ward tend to have large receive scores considering their overall activity. The results are also presented in Appendix \ref{lists}. 

Fig.(\ref{plots}) shows the expected results of performing various shufflings, we can think of the $y$-axis in these plots as a measure of how much the score of each individual depends on temporal properties, and the $x$-axis for structural properties. We see that the outlier from the Enron data-set is, remarkably, unremarkable regarding both of these measures and neither alone can explain their high broadcast score (time-shuffled rank $86$ and target-shuffled rank $104$, both lower than the actual broadcast rank of $50$). However, the fact that both shuffled ranks are higher than the degree rank suggests that the individual in question is sending emails very economically, i.e. sending to the most efficient recipients, and choosing the optimal moments in time to send. From this example it appears that the contributions from both factors add to the overall broadcast score.

An alternative interpretation is that the individual in question was feeding information into the network which was consequently being disseminated in a way that inflates their broadcast score (although similar results are not found for the CEOs who we would expect to be influential in the same way). The individual in question was a lobbyist for the corporation, after a very brief investigation we did not determine a particular reason why they should be significantly influential.

From Fig.(\ref{hospital_plot2}) it is apparent that shuffling the time column can cause large changes to the receive rank of a participant whereas the source-shuffling appears to be less effective. This is because the temporal activity of the participants deviates significantly from a Bernoulli process (that is assumed in the time-independent model). More specifically, nodes exist which are inactive towards the beginning of the sampling period but have a lot of activity at later time-steps. The receive score of these nodes is amplified by the exponential increase over time that is indicated by the very simple example in Section \ref{simple_time}. Those which are active early on in the sampling period but have little or no activity at later times will have lower receive scores. When such effects dominate the outcome the effect of time-shuffling is significant. 

While we do not discuss here the broadcast scores for the Hospital data, or the receive scores for the Enron email data, the results can be seen in Fig.(\ref{box_appendix}), Fig.(\ref{plots_2_appendix}) and Fig.(\ref{plots_appendix}).

\begin{figure*}[t]
    \centering
		   \subfloat[Enron email directory\label{enron_plot2}]{
        \includegraphics[width=0.46\textwidth]{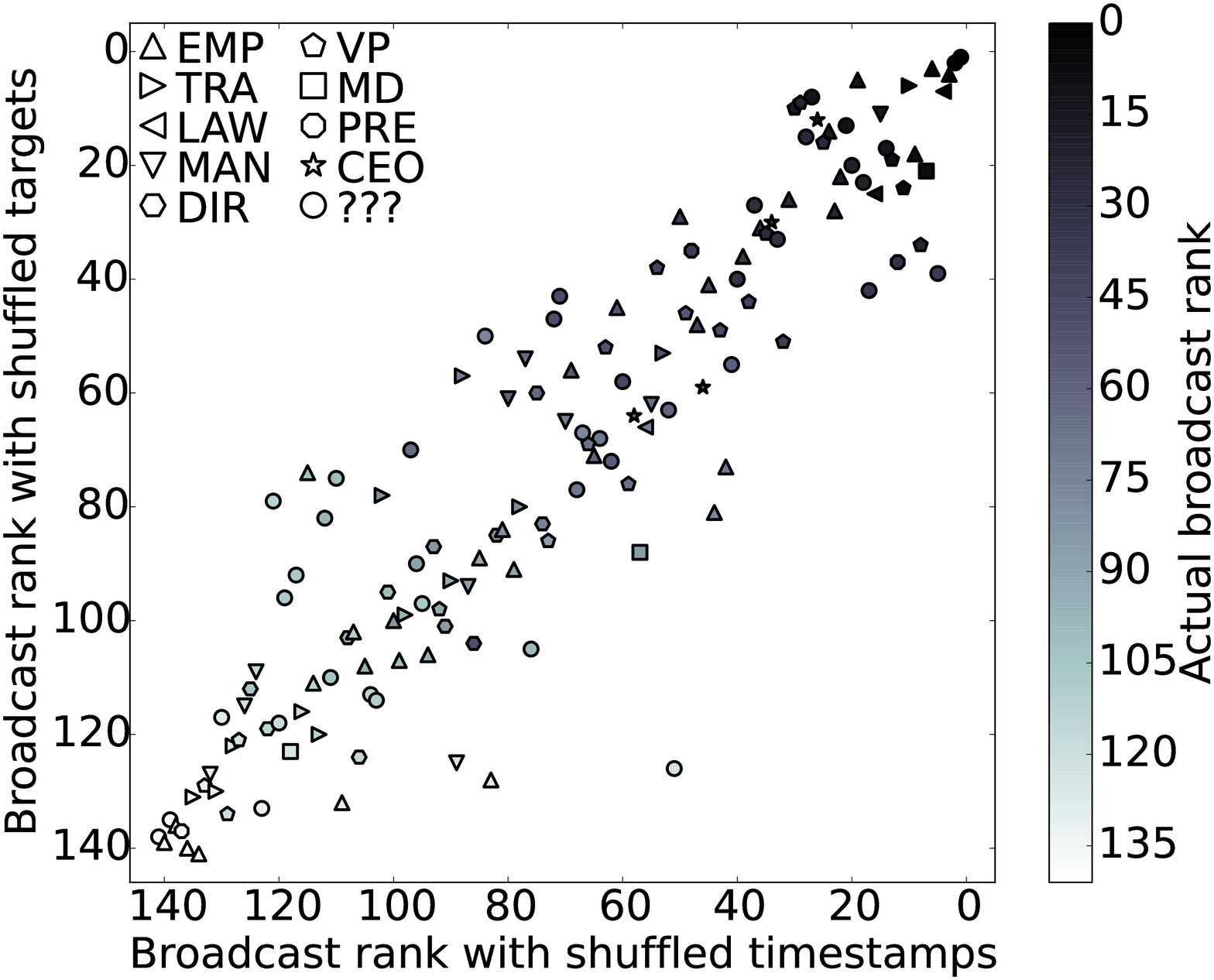}	
				}\quad
				\subfloat[Hospital contact network\label{hospital_plot2}]{
        \includegraphics[width=0.46\textwidth]{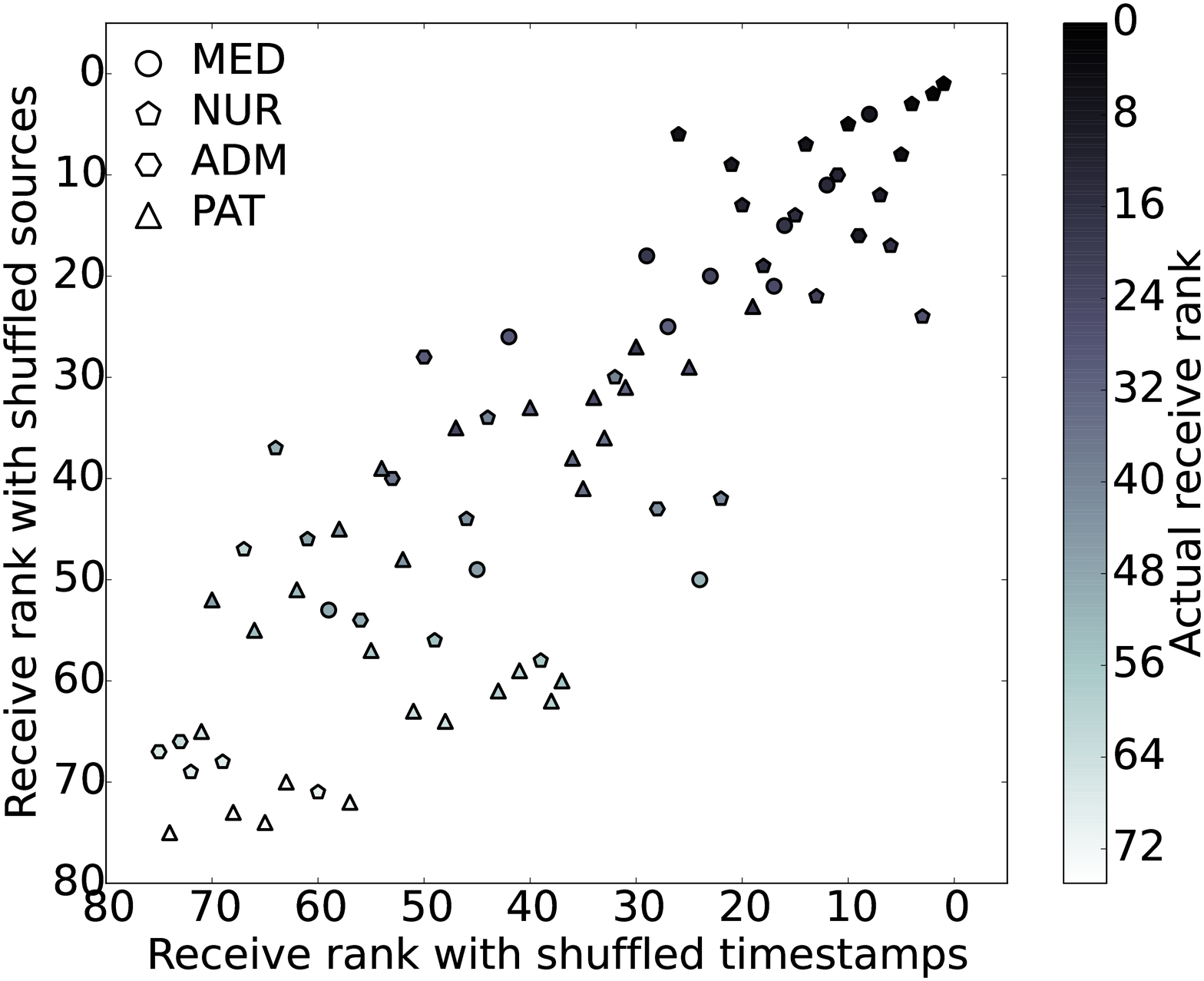}
	}
    \caption{As demonstrated in Fig.(\ref{databases}). On the $y$-axis we show the ranking of each node according to expectation of the broadcast score (left, computed using Eq.\eqref{time_formula_b}) and receive score (right, computed with Eq.\eqref{time_formula}) for the expected outcomes of the source (left) and target (right) shuffled networks (with $\alpha=0.01$ and $\alpha=0.005$ respectively). The $x$-axes show the expected scores for a time-shuffled network computed with Eqs.\eqref{W_sol_b} and \eqref{W_sol_c}. The actual broadcast score computed with Eqs. \eqref{bc_compute} and \eqref{rc_compute} is shown by the darkness of the markers. Different roles are indicated by the marker shapes, the abbreviations are explained in Section \ref{data}. These results are also presented in Appendix \ref{lists}.
\label{plots}}
\end{figure*}

\section{Discussion}
\label{discussion}
As data-driven industries increasingly find value in targeting the most central, most influential, individuals, it is important to scrutinize the methods and tools that network science is promoting. The idea that there is one magic formula which can produce a meaningful result regardless of the system in question is firstly, wrong, and secondly, a counter-productive way of thinking. Here we have scrutinized the dynamic communicability metrics and found that temporal variation can have a stronger effect in some systems, like the hospital ward, than in others, like Enron. We have found efficient shortcut formulae to quantify the temporal component by randomizing the structural factors and likewise quantify the structural component by randomizing the temporal factors. Those who have data and wish to analyze dynamic communicability should use these methods to add more dimensions, and more depth, to their analysis.

When we look at the simple example of Fig.(\ref{big_fig}), we can compute the broadcast scores and find that node $A$ is ranked number one. We can then ask why $A$ is the most influential broadcaster and find that it is not because it was the most active ($C$ was in fact the most active), but because of a complex interplay of temporal and structural factors; $A$ was the first to communicate, and importantly, one of those early edges was received by $C$ who was subsequently the most active node. Looking at large data-sets it is tedious to try to deconstruct every sequence of contacts that caused each individual to achieve its score. Instead, we have introduced meaningful statistics, i.e. the results of shuffling, that provide insight into the interplay of temporal and structural factors.

Several specific phenomena are commonly found in social systems whose effects are nullified by the shuffling process. The structure (or topology) of complex networks has been extensively studied and many elements have repeatedly been found across different systems \cite{newman2003structure}. Degree heterogeneity is one such topological feature which is not nullified by shuffling. On the other hand, features like community structure, assortativity and clustering are likely to play a significant role in determining the communicability in most applications of this work \cite{clauset2004finding,barclay2014positive,malik2013role}. Similarly, the non-shuffled data is likely to exhibit certain temporal features. Recent studies of communication data, similar to the Enron data-set, show that activity generally occurs in bursts \cite{goh2008burstiness}. Others focus on the effect of circadian cycles which are likely to occur in the hospital ward data \cite{jo2012circadian}. 

Clustering and burstiness both increase the number of walks which revisit nodes. For most contagion processes these walks would not be permissible since, for example, many diseases can only be contracted once, similarly a piece of information can only be attained once (this is possibly the reason why bursty networks have been shown to slow the spread of information compared to their temporally shuffled equivalent \cite{karsai2011small}). This remains a fundamental problem of the dynamic communicability metric which should not be overlooked.

Another issue that ought to be considered when using the dynamic communicability metrics is the effect of a bounded sampling window. Take for example the simple example of Fig.(\ref{big_fig}). Here $A$ has the highest broadcast score because it is the first node to create outgoing edges. Had we observed the system just one time-step earlier we might have found one or more edges from $C$ to $A$, thus making $C$ the highest ranked broadcaster above $A$. This is a general issue; our analytical results tell us that the earlier interactions contribute exponentially more than those which occur later; therefore the first node involved in the first recorded interaction will, by chance, receive an unduly high broadcast score. In the case of the receive score, interactions that occur late in the sample inflate the score of the involved nodes. The advancement of dynamic communicability presented in \cite{grindrod2013matrix}, that assumes infectiousness decays in the time between interactions, may mitigate these problems to some extent. We conclude this paper by suggesting two possible alternative solutions:
\subsection{Control for temporal variation}
Eq.\eqref{time_formula} gives the expectation of the receive score based on temporal variation. It can therefore be considered as a control to compare to the actual score. Further, we suggest that a normalized version of the receive score would be a more appropriate measure to compare individuals in the same network. The normalized version is the ratio of the actual score, computed using Eq.\eqref{rc_compute}, and its expectation, computed using Eq.\eqref{time_formula}. 
\subsection{Remove temporal variation}
Alternatively, we ignore temporal variation altogether; in many circumstances this is sensible since the temporal variation over the duration of the sample is not usually expected to be the same in the future (unless perhaps it is driven by a cyclic process). Without knowledge of when each future interaction will occur, the Bernoulli process used in the time-independent model is a suitable choice. In such a case, the past data provides an estimate of how active each node will be, but the timing of their interactions remains random. The matrix exponential in Eqs.\eqref{W_sol_b} and \eqref{W_sol_c}, can be computed very efficiently to give these approximations to the receive score and broadcast score. Incidentally, the matrix exponential has previously been proposed as a centrality measure \cite{estrada2008communicability,benzi2013total}.

\section*{Acknowledgements}
We thank Georgios Giasemidis for helpful discussions at early stages of the project. We are grateful to Shweta Bansal for helpful comments regarding the structure and presentation of the manuscript and to Isabel Chen for feedback in the late stages. E.R.C was funded in part by RCUK Digital Economy programme via EPSRC Grant EP/G065802/1 `The Horizon Hub' and in part by NSF Grant No. $1414296$ as part of the joint NSF-NIH-USDA Ecology and Evolution of Infectious Diseases program.
\end{multicols}

\small
\bibliography{bibfile}
\bibliographystyle{ieeetr}
\normalsize

\newpage
\begin{appendices}
\begin{multicols}{2}
\begin{figure*}[t]
\centering
\subfloat[Hospital results \label{hospital_raw_appendix}]{
		\includegraphics[height=6cm]{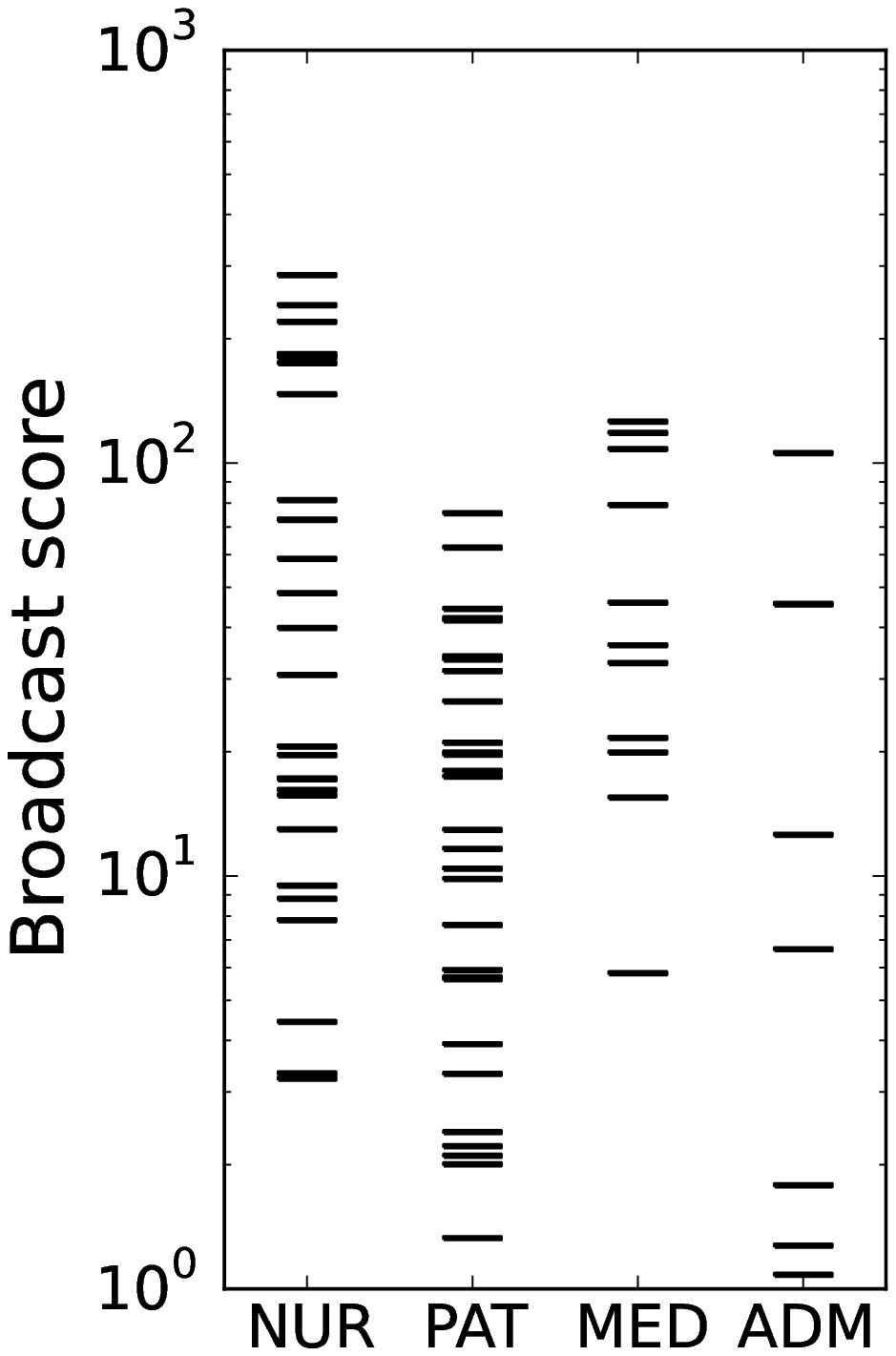}}\qquad
    \subfloat[Enron email results \label{enron_raw_appendix}]{
		\includegraphics[height=6cm]{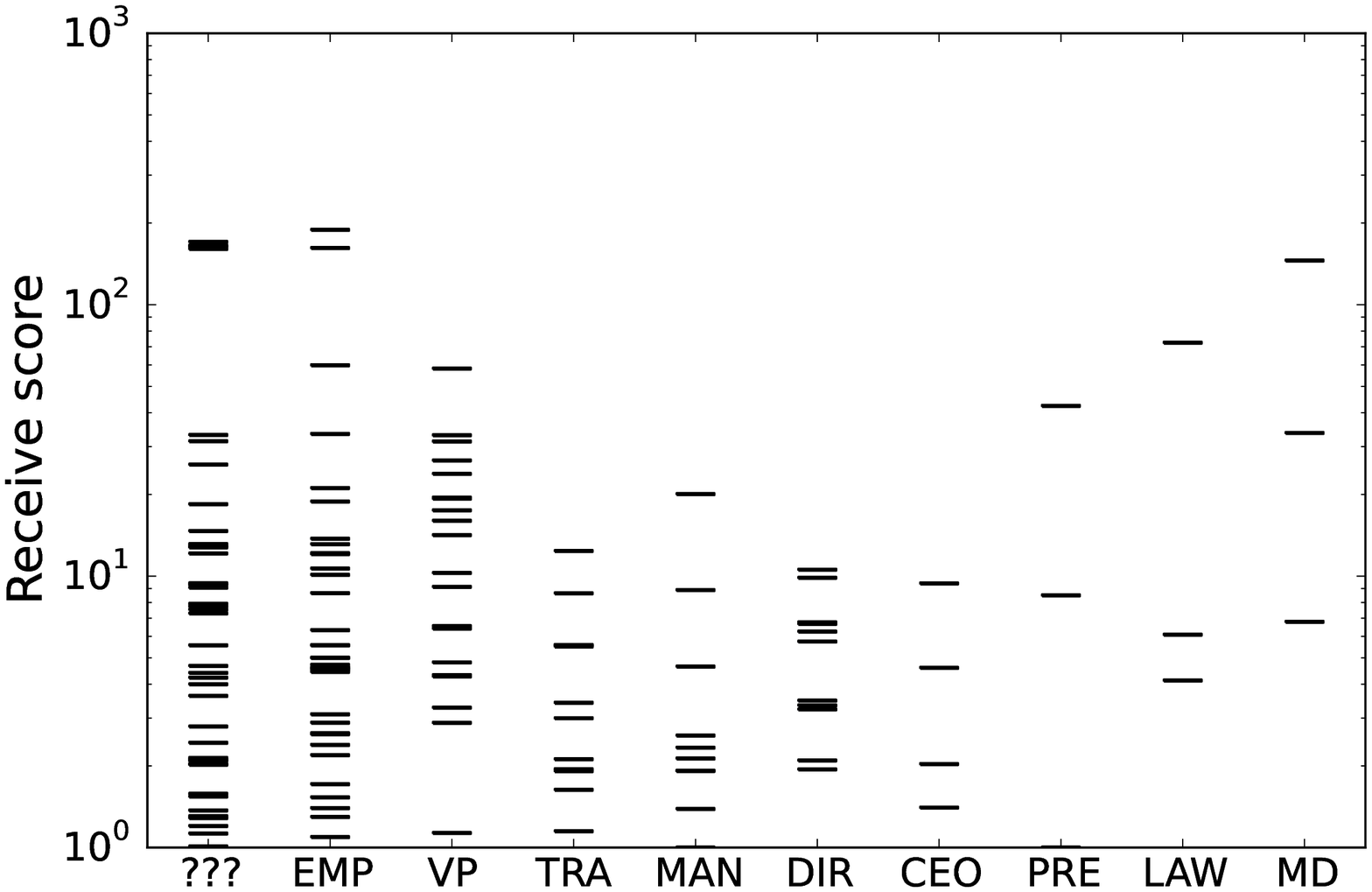}}
    \caption{Each marker (short horizontal line) represents an Enron employee or participant in the Sociopatterns experiment. The broadcast score, computed by Eq.\eqref{bc_compute}, is displayed on the $y$-axis in (\ref{hospital_raw_appendix}) and similarly the receive score in (\ref{enron_raw_appendix}) from Eq.\eqref{rc_compute} with $\alpha=0.005$ and $\alpha=0.01$ respectively. In both the employees (or participants) are divided into distinct categories shown along the $x$-axis. Abbreviations are given in Section \ref{data}.}
		\label{box_appendix}
\end{figure*}

\section{Modeling}
\subsection{Heterogeneous ``send'' and ``receive'' model}
\label{sr_sol}
\emph{The Model:}
\begin{quotation}
\noindent
In any given time-step, the probability that $i$ has an out going edge is $s_{i}$, the probability that it has an incoming edge is $r_{i}$.
\end{quotation}
Making no further assumptions about who communicates with whom, letting $\mathbf{r}$ and $\mathbf{s}$ both be column vectors we have the general stochastic model with
\begin{equation}
P=\mathbf{s}\mathbf{r}^{T}=
 \begin{pmatrix}
  s_{1}r_{1} & s_{1}r_{2}& \cdots & s_{1}r_{N} \\
  s_{2}r_{1} & s_{2}r_{2} & \cdots & s_{2}r_{N} \\
  \vdots  & \vdots  & \ddots & \vdots  \\
  s_{N}r_{1} & s_{N}r_{2} & \cdots & s_{N}r_{N}
 \end{pmatrix}.
\end{equation}
There are at least two ways to find the expectation of broadcast and receive scores for this model. It is possible to write down an expression for the $P^{k}$ which can then be substituted into Eq.(\ref{powers}). An alternative method is to solve Eq.(\ref{diff_c}) directly. First we express Eq.(\ref{diff_c}) in terms of our new variables:
\begin{equation}
\label{c_diff_rs}
\frac{\partial \mathbf{\hat{c}}(t)}{\partial t}=\alpha \mathbf{\hat{c}}(t)\mathbf{s}\mathbf{r}^{T} .
\end{equation}
Multiplying both sides on the right by $s$ gives
\begin{equation}
\frac{\partial \mathbf{\hat{c}}(t)\mathbf{s}}{\partial t}=\alpha \mathbf{\hat{c}}(t)\mathbf{s}\mathbf{r}^{T}\mathbf{s}.
\end{equation}
which is a differential equation describing the time-evolution of $\mathbf{\hat{c}}(t)\mathbf{s}$, a scalar variable. This has the solution
\begin{equation}
\mathbf{\hat{c}}(t)\mathbf{s}=e^{\alpha[\mathbf{r}^{T}\mathbf{s}]t}.
\end{equation}
Substituting the result back into Eq.\eqref{c_diff_rs} we get
\begin{equation}
\frac{\partial \mathbf{\hat{c}}(t)}{\partial t}=\alpha \mathbf{r}^{T}e^{\alpha[\mathbf{r}^{T}\mathbf{s}]t}.
\end{equation}
Which has the solution
\begin{equation}
\mathbf{\hat{c}}(t)=\mathbf{1}+\left(\frac{e^{\alpha[\mathbf{r}^{T}\mathbf{s}]t}-1}{\mathbf{r}^{T}\mathbf{s}}\right)\mathbf{r}^{T}.
\end{equation}
In a similar way one can show that the expectation of the broadcast score is
\begin{equation}
\mathbf{\hat{b}}(t)=\mathbf{1}^{T}+\left(\frac{e^{\alpha[\mathbf{r}^{T}\mathbf{s}]t}-1}{\mathbf{r}^{T}\mathbf{s}}\right)\mathbf{s}.
\end{equation}

\subsection{Time-dependent $P$ matrix}
\label{t_model}
\emph{The Model:}
\begin{quotation}
\noindent
At time $\tau$, the probability that $i$ has an out going edge is $s_{i}$, the probability that it has an incoming edge is $r_{i}(\tau)$
\end{quotation}
Eq.(\ref{diff_c}) now becomes
\begin{equation}
\label{diff_s}
\frac{\partial \mathbf{\hat{c}}(t)}{\partial t}=\alpha \mathbf{\hat{c}}(t)\mathbf{s}\mathbf{r}(t_{0}+t)^{T}.
\end{equation}
Multiplying both sides on the right by $\mathbf{s}$ we get
\begin{equation}
\begin{split}
\frac{\partial \mathbf{\hat{c}}(t)}{\partial t}\mathbf{s}&=\alpha \mathbf{\hat{c}}(t) \mathbf{s}\mathbf{r}(t_{0}+t)^{T}\mathbf{s}\\
\frac{\partial \mathbf{\hat{c}}(t)\mathbf{s}}{\partial t}&=\alpha [\mathbf{\hat{c}}(t)\mathbf{s}][\mathbf{r}(t_{0}+t)^{T}\mathbf{s}].
\end{split}
\end{equation}
This equation now only includes scalar functions of $t$ so we can solve to get
\begin{equation}
\mathbf{\hat{c}}(t)\mathbf{s}=\mathbf{1}\mathbf{s}\exp\left(\alpha \int_{0}^{t}\mathbf{s}\mathbf{r}(t_{0}+t')^{T}dt' \right).
\end{equation}
Substituting this back into Eq.\eqref{diff_s} we have
\begin{equation}
\label{diff_srt}
\frac{\partial \mathbf{\hat{c}}(t)}{\partial t}=\alpha \mathbf{r}(t_{0}+t)^{T}\exp\left(\alpha \int_{0}^{t}\mathbf{s}\mathbf{r}(t_{0}+t')^{T}dt' \right)
\end{equation}
Since $\mathbf{1}s=1$.
\subsection{Simple time-dependent example}
\label{simple_time2}
\emph{The model:}
\begin{quotation}
\noindent
At time $\tau_{i}$ person $i$ is on the receiving end of $r_{i}$ edges. As before, the number of outgoing edges is determined by a time-independent probability $s_{i}$.
\end{quotation}
Clearly, after $N$ iterations the process will end so we use $t_{0}=0$ and $t_{\text{end}}=N$ as the initial and final conditions respectively. To find the broadcast score of a node $i$ we solve Eq.\eqref{diff_srt} with 
\begin{equation}
r_{i}(\tau)=r_{i}\delta(\tau-\tau_{i})
\end{equation}
where $r_{i}$ is a scalar and $\delta$ is the Dirac delta. The justification for this version of $r_{i}(\tau)$ is that the expected number of messages sent by $i$ over some time-interval will be $r_{i}$ if the time interval includes $\tau_{i}$. Without loss of generality we can say $\tau_{i}=i$ meaning that node $1$ sends first, then node $2$ and so on. First we focus on expressing $\mathbf{s}\int_{0}^{t}\mathbf{r}(t_{0}+t')^{T} d t'$ in a simpler form. Since
\begin{equation}
\int_{0}^{t}r_{j}\delta(t'-j)dt'=\left\{
   \begin{aligned}
     & r_{j}& \quad \text{if}\quad j\leq t\\
     &0 & \quad \text{if}\quad j\geq t
   \end{aligned}\right.
\end{equation}
(This result derives from the fact that the integral of the Dirac delta between $-\infty$ and $t$ is the Heaviside step function $H(t)$.) we have
\begin{equation}
\begin{split}
\mathbf{s}\int_{0}^{t}\mathbf{r}(t_{0}+t')^{T} d t'&=\sum_{j=0}^{N} s_{j}\int_{0}^{t}r_{j}\delta(t'-j) d t'\\
&=\sum_{j=0}^{t}r_{j}s_{j}.
\end{split}
\end{equation}
Substituting this into Eq.\eqref{diff_srt} then integrating over the whole sample gives
\begin{equation}
\left[\hat{c}_{i}(t')\right]_{0}^{N}=\int_{0}^{N}\alpha r_{i}\delta(t'-i)\exp\left(\alpha\sum_{j=0}^{i}r_{j}s_{j}\right)dt'.
\end{equation}
The integral is solved by the translation property of the Dirac delta and we have
\begin{equation}
\hat{c}_{i}(N)=1+\alpha r_{i}\exp\left(\alpha\sum_{j=0}^{i}r_{j}s_{j}\right).
\end{equation}

\subsection{Incorporating empirical data}
\label{source_shuffled_derivation}
\emph{The model:}
\begin{quotation}
\noindent
Let $K_{i}$ be the set of edges for which $i$ is the target node, and $\tau^{(k)}$ be the time at which edge $k$ was present. As before, $s_{i}$ is the time-independent probability for $i$ to be the source of an edge.
\end{quotation}
We achieve this by choosing
\begin{equation}
r_{i}(t)=\sum_{k\in K_{i}}\delta(t-\tau_{i}^{(k)}).
\end{equation}
We can choose the set $K_{i}$ and the corresponding $\tau^{(k)}$ in a way that recreates exactly what is observed in the target and time columns of an empirical temporal edge-list. We introduce $R_{i}(\tau)$, the number of messages sent by $i$ between time $t_{0}$ and time $\tau$, this is expressed
\begin{equation}
R_{i}(\tau)=\int_{t_{0}}^{\tau}r_{i}(t')dt',
\end{equation}
giving
\begin{equation}
R_{i}(t_{0}+t)=\int_{0}^{t}r_{i}(t_{0}+t')dt',
\end{equation}
and therefore Eq.\eqref{diff_srt} can be expressed
\begin{equation}
\frac{\partial \mathbf{\hat{c}}(t)}{\partial t}=\alpha \mathbf{r}(t_{0}+t)^{T}e^{\alpha \sum_{j=1}^{N}s_{j}R_{j}(t_{0}+t)}.
\end{equation}
Integrating over the entire duration of the sample gives
\begin{equation}
\begin{split}
&\left[\hat{c}_{i}(t')\right]_{0}^{t_{\text{end}}-t_{0}}\\
&=\alpha\int_{0}^{t_{\text{end}}-t_{0}}\left[\sum_{k \in K_{i}}\delta(t_{0}+t'-\tau^{(k)})\right]\exp\left(\alpha\sum_{j=1}^{N}s_{j}R_{j}(t_{0}+t')\right)dt'\\
&=\alpha\sum_{k \in K_{i}}\int_{-\infty}^{\infty}\delta[t_{0}+t'-\tau^{(k)}]\exp\left(\alpha\sum_{j=1}^{N}s_{j}R_{j}(t_{0}+t')\right)dt'.
\end{split}
\end{equation}
Finally, using the translation property of the Dirac delta function we have
\begin{equation}
\hat{c}_{i}=1+\alpha\sum_{k\in K_{i}}\exp\left(\alpha\sum_{j\in N}s_{j}R_{j}(\tau_{i}^{(k)})\right).
\end{equation}
\begin{figure*}[t!]
    \centering
				\subfloat[Hospital contact network\label{hospital_plot}]{
        \includegraphics[width=0.46\textwidth]{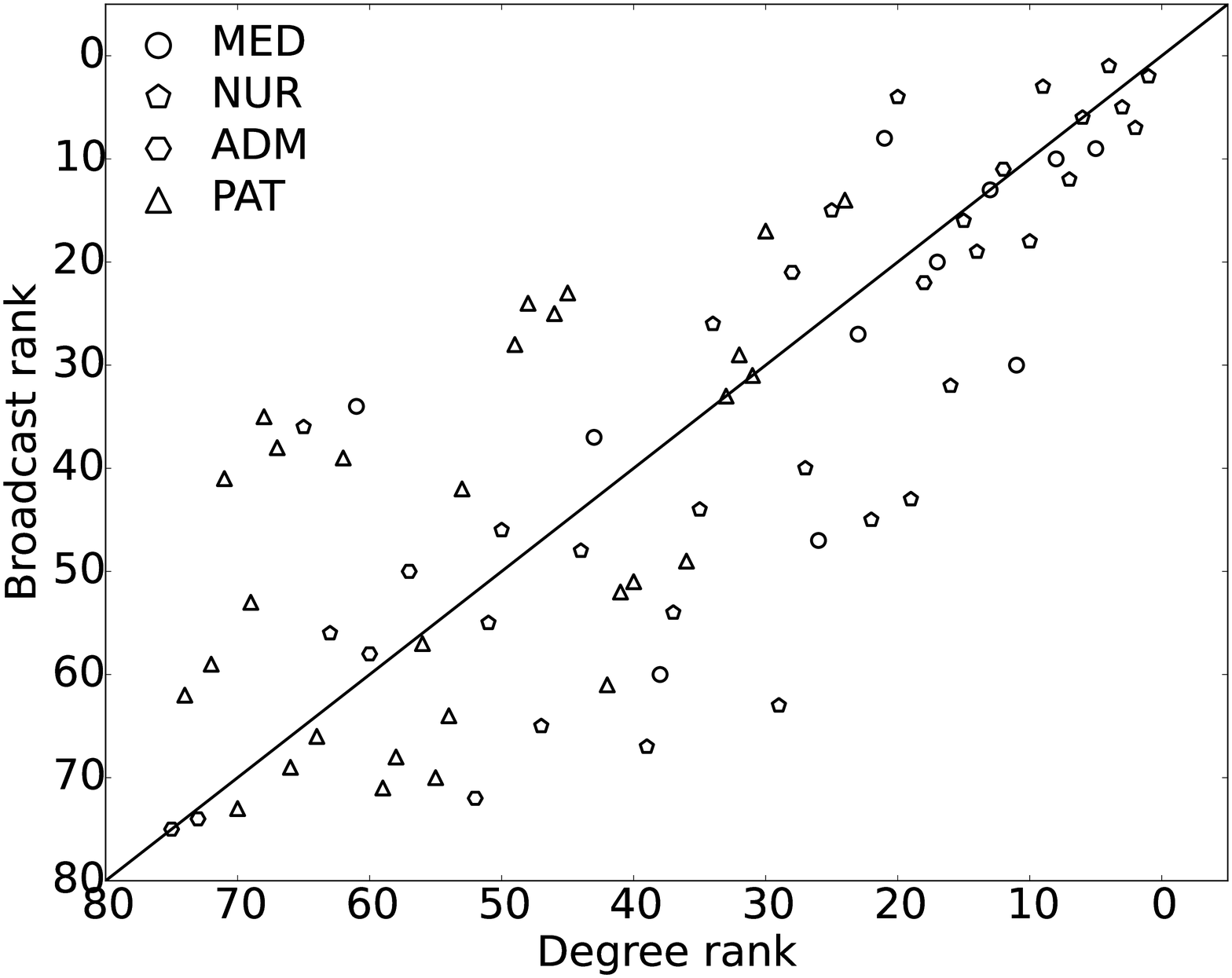}}\quad
    \subfloat[Enron email directory\label{enron_plot}]{
        \includegraphics[width=0.46\textwidth]{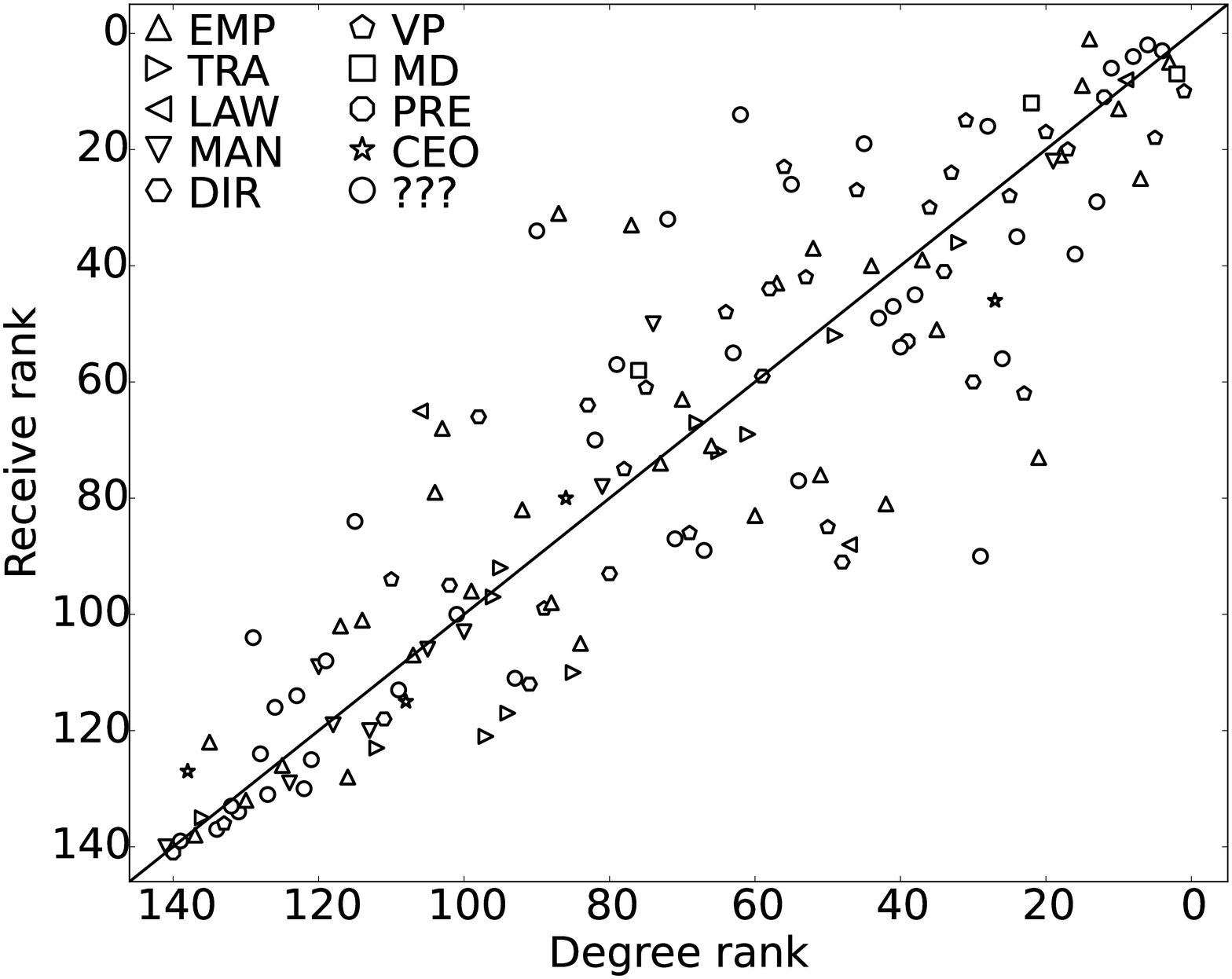}
				}
    \caption{The rank according to broadcast score (left, computed by Eq.\eqref{bc_compute}) and receive score (right, computed by Eq.\eqref{rc_compute}), with $\alpha=0.005$ and $\alpha=0.01$ respectively, plotted against the out-degree (left) and in-degree (right). Each individual in the network is represented by a data point, their classification is given by their shape. The abbreviations in the legend are explained in Section \ref{data}. The one-to-one line is plotted as a visual aid to partition the nodes into two groups; those which have higher than expected scores (top left), and those who have lower than expected (bottom right).}
\label{plots_2_appendix}
\end{figure*}

\begin{figure*}[t!]
    \centering
			 \subfloat[Hospital contact network\label{hospital_plot2_appendix}]{
        \includegraphics[width=0.46\textwidth]{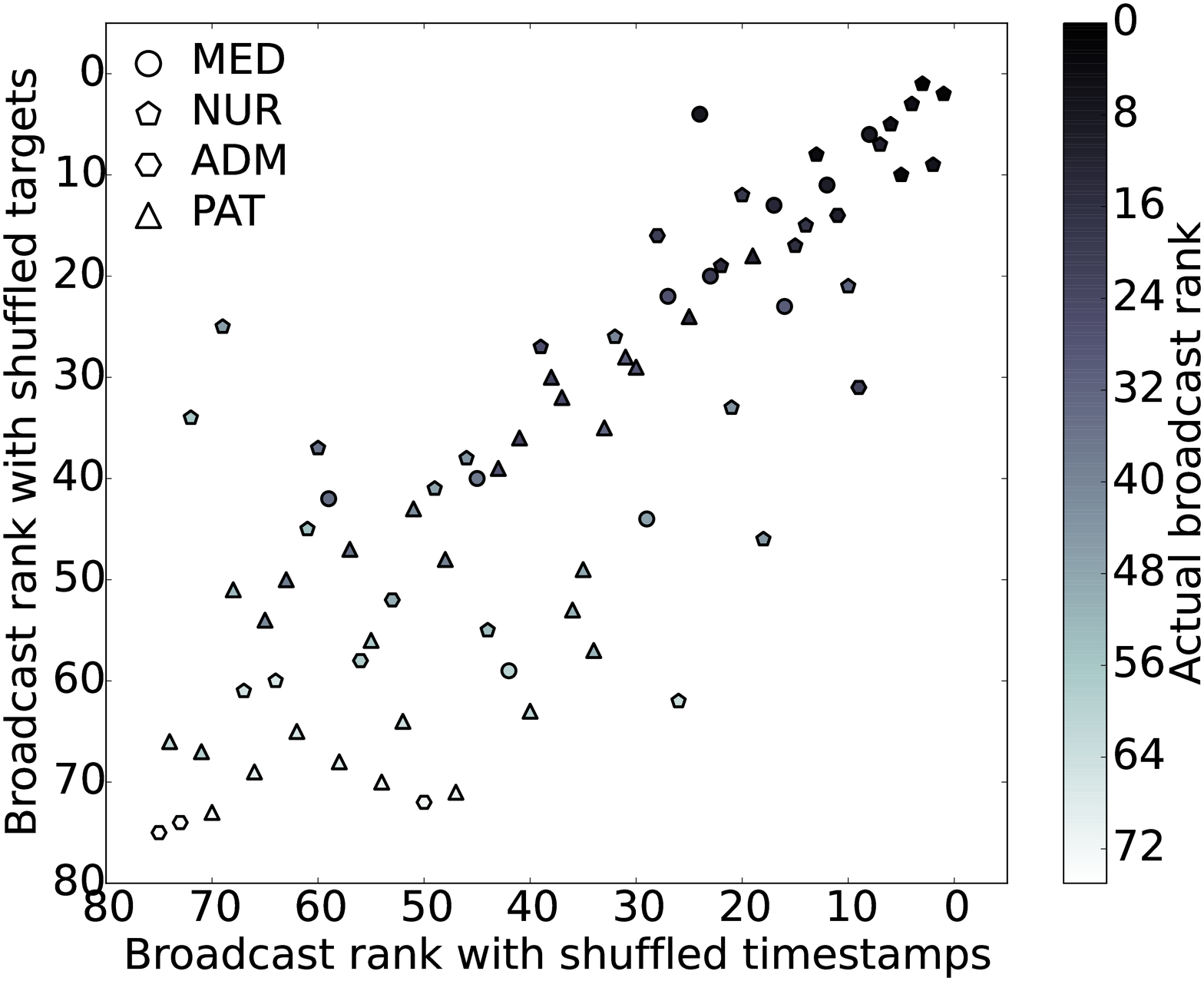}}\quad
		   \subfloat[Enron email directory\label{enron_plot2_appendix}]{
        \includegraphics[width=0.46\textwidth]{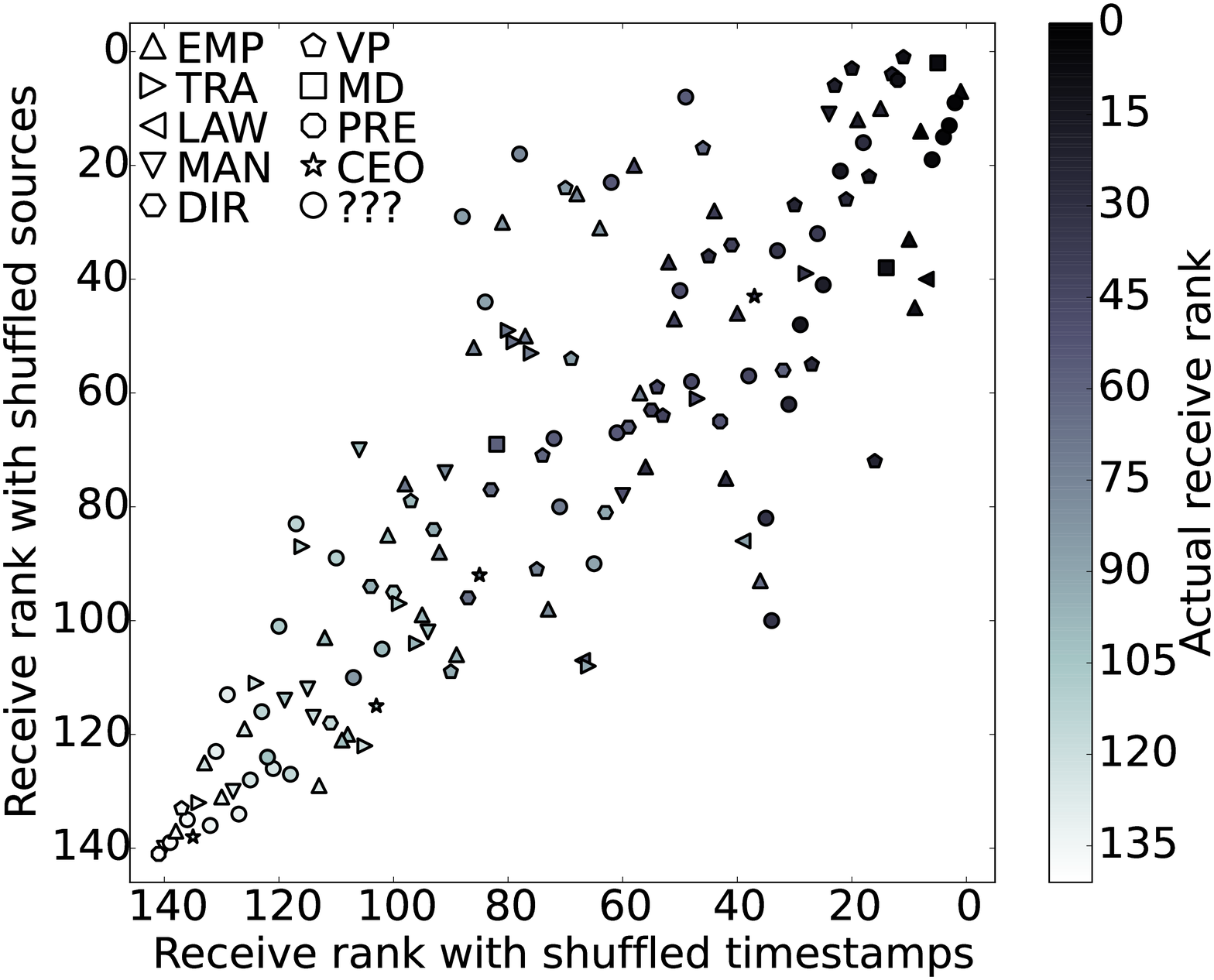}	
				}
    \caption{As demonstrated in Fig.(\ref{databases}). On the $y$-axis we show the ranking of each node according to expectation of the broadcast score (left, computed using Eq.\eqref{time_formula_b}) and receive score (right, computed with Eq.\eqref{time_formula}) for the expected outcomes of the source (left) and target (right) shuffled networks (with $\alpha=0.005$ and $\alpha=0.01$ respectively). The $x$-axes show the expected scores for a time-shuffled network computed with Eqs.\eqref{W_sol_b} and \eqref{W_sol_c}. The actual broadcast score computed with Eqs. \eqref{bc_compute} and \eqref{rc_compute} is shown by the darkness of the markers. Different roles are indicated by the marker shapes, the abbreviations are explained in Section \ref{data}.
\label{plots_appendix}}
\end{figure*}
\end{multicols}

\newpage
\newgeometry{bottom=2cm,top=1cm}
\section{Rankings}
\label{lists}
\tiny
\subsection{Sociopatterns hospital ward receive-rank}
\begin{longtable*}{l|l|l|l|l}
Rank	& None 	& Time-shuffled	& Source-shuffled & Time and Source \\
\hline
1 & 1115 (NUR) &1115 (NUR) &1115 (NUR) &1115 (NUR) \\ 

2 & 1210 (NUR) &1210 (NUR) &1210 (NUR) &1210 (NUR) \\ 

3 & 1190 (NUR) &1207 (NUR) &1295 (NUR) &1295 (NUR) \\ 

4 & 1295 (NUR) &1295 (NUR) &1157 (MED) &1207 (NUR) \\ 

5 & 1109 (NUR) &1109 (NUR) &1190 (NUR) &1157 (MED) \\ 

6 & 1629 (NUR) &1164 (NUR) &1629 (NUR) &1164 (NUR) \\ 

7 & 1149 (NUR) &1193 (NUR) &1149 (NUR) &1193 (NUR) \\ 

8 & 1157 (MED) &1157 (MED) &1109 (NUR) &1144 (MED) \\ 

9 & 1205 (NUR) &1658 (ADM) &1205 (NUR) &1109 (NUR) \\ 

10 & 1658 (ADM) &1190 (NUR) &1098 (ADM) &1149 (NUR) \\ 

11 & 1193 (NUR) &1098 (ADM) &1144 (MED) &1221 (MED) \\ 

12 & 1196 (NUR) &1144 (MED) &1193 (NUR) &1098 (ADM) \\ 

13 & 1098 (ADM) &1114 (NUR) &1196 (NUR) &1159 (MED) \\ 

14 & 1144 (MED) &1149 (NUR) &1181 (NUR) &1196 (NUR) \\ 

15 & 1181 (NUR) &1181 (NUR) &1221 (MED) &1181 (NUR) \\ 

16 & 1625 (NUR) &1221 (MED) &1658 (ADM) &1190 (NUR) \\ 

17 & 1164 (NUR) &1159 (MED) &1164 (NUR) &1260 (MED) \\ 

18 & 1221 (MED) &1625 (NUR) &1130 (MED) &1658 (ADM) \\ 

19 & 1130 (MED) &1365 (PAT) &1625 (NUR) &1205 (NUR) \\ 

20 & 1365 (PAT) &1196 (NUR) &1260 (MED) &1114 (NUR) \\ 

21 & 1383 (PAT) &1205 (NUR) &1159 (MED) &1191 (MED) \\ 

22 & 1114 (NUR) &1245 (NUR) &1114 (NUR) &1625 (NUR) \\ 

23 & 1260 (MED) &1260 (MED) &1365 (PAT) &1148 (MED) \\ 

24 & 1547 (PAT) &1191 (MED) &1207 (NUR) &1365 (PAT) \\ 

25 & 1159 (MED) &1378 (PAT) &1148 (MED) &1245 (NUR) \\ 

26 & 1702 (PAT) &1629 (NUR) &1660 (MED) &1130 (MED) \\ 

27 & 1207 (NUR) &1148 (MED) &1383 (PAT) &1202 (NUR) \\ 

28 & 1378 (PAT) &1179 (ADM) &1671 (ADM) &1179 (ADM) \\ 

29 & 1660 (MED) &1130 (MED) &1378 (PAT) &1629 (NUR) \\ 

30 & 1671 (ADM) &1383 (PAT) &1202 (NUR) &1378 (PAT) \\ 

31 & 1148 (MED) &1352 (PAT) &1352 (PAT) &1352 (PAT) \\ 

32 & 1401 (PAT) &1202 (NUR) &1702 (PAT) &1383 (PAT) \\ 

33 & 1352 (PAT) &1391 (PAT) &1401 (PAT) &1391 (PAT) \\ 

34 & 1307 (PAT) &1702 (PAT) &1142 (NUR) &1105 (NUR) \\ 

35 & 1362 (PAT) &1362 (PAT) &1547 (PAT) &1108 (NUR) \\ 

36 & 1391 (PAT) &1307 (PAT) &1391 (PAT) &1362 (PAT) \\ 

37 & 1232 (ADM) &1374 (PAT) &1485 (NUR) &1142 (NUR) \\ 

38 & 1469 (PAT) &1393 (PAT) &1307 (PAT) &1660 (MED) \\ 

39 & 1202 (NUR) &1105 (NUR) &1469 (PAT) &1485 (NUR) \\ 

40 & 1142 (NUR) &1401 (PAT) &1232 (ADM) &1307 (PAT) \\ 

41 & 1245 (NUR) &1363 (PAT) &1362 (PAT) &1702 (PAT) \\ 

42 & 1179 (ADM) &1660 (MED) &1245 (NUR) &1401 (PAT) \\ 

43 & 1108 (NUR) &1395 (PAT) &1179 (ADM) &1168 (MED) \\ 

44 & 1701 (PAT) &1142 (NUR) &1108 (NUR) &1100 (NUR) \\ 

45 & 1460 (PAT) &1168 (MED) &1460 (PAT) &1393 (PAT) \\ 

46 & 1168 (MED) &1108 (NUR) &1261 (NUR) &1374 (PAT) \\ 

47 & 1784 (PAT) &1547 (PAT) &1613 (NUR) &1613 (NUR) \\ 

48 & 1261 (NUR) &1320 (PAT) &1701 (PAT) &1363 (PAT) \\ 

49 & 1152 (MED) &1100 (NUR) &1168 (MED) &1395 (PAT) \\ 

50 & 1209 (ADM) &1671 (ADM) &1191 (MED) &1246 (NUR) \\ 

51 & 1485 (NUR) &1327 (PAT) &1769 (PAT) &1261 (NUR) \\ 

52 & 1191 (MED) &1701 (PAT) &1784 (PAT) &1671 (ADM) \\ 

53 & 1769 (PAT) &1232 (ADM) &1152 (MED) &1327 (PAT) \\ 

54 & 1416 (PAT) &1469 (PAT) &1209 (ADM) &1701 (PAT) \\ 

55 & 1100 (NUR) &1385 (PAT) &1416 (PAT) &1547 (PAT) \\ 

56 & 1374 (PAT) &1209 (ADM) &1100 (NUR) &1385 (PAT) \\ 

57 & 1105 (NUR) &1399 (PAT) &1385 (PAT) &1232 (ADM) \\ 

58 & 1385 (PAT) &1460 (PAT) &1105 (NUR) &1460 (PAT) \\ 

59 & 1395 (PAT) &1152 (MED) &1363 (PAT) &1469 (PAT) \\ 

60 & 1393 (PAT) &1116 (NUR) &1374 (PAT) &1209 (ADM) \\ 

61 & 1363 (PAT) &1261 (NUR) &1395 (PAT) &1152 (MED) \\ 

62 & 1613 (NUR) &1769 (PAT) &1393 (PAT) &1320 (PAT) \\ 

63 & 1535 (ADM) &1377 (PAT) &1327 (PAT) &1238 (NUR) \\ 

64 & 1327 (PAT) &1485 (NUR) &1320 (PAT) &1769 (PAT) \\ 

65 & 1320 (PAT) &1323 (PAT) &1373 (PAT) &1116 (NUR) \\ 

66 & 1373 (PAT) &1416 (PAT) &1535 (ADM) &1416 (PAT) \\ 

67 & 1525 (ADM) &1613 (NUR) &1525 (ADM) &1377 (PAT) \\ 

68 & 1246 (NUR) &1305 (PAT) &1246 (NUR) &1399 (PAT) \\ 

69 & 1238 (NUR) &1246 (NUR) &1238 (NUR) &1305 (PAT) \\ 

70 & 1116 (NUR) &1784 (PAT) &1377 (PAT) &1784 (PAT) \\ 

71 & 1399 (PAT) &1373 (PAT) &1116 (NUR) &1323 (PAT) \\ 

72 & 1377 (PAT) &1238 (NUR) &1399 (PAT) &1373 (PAT) \\ 

73 & 1305 (PAT) &1535 (ADM) &1305 (PAT) &1535 (ADM) \\ 

74 & 1323 (PAT) &1332 (PAT) &1323 (PAT) &1332 (PAT) \\ 

75 & 1332 (PAT) &1525 (ADM) &1332 (PAT) &1525 (ADM) \\ 
\end{longtable*}

\subsection{Enron email broadcast rank}
\begin{longtable*}{l|l|l|l|l}
Rank	& None 	& Time-shuffled	& Target-shuffled & Time and Target \\
\hline
1 & tana.jones (???) & tana.jones (???) & tana.jones (???) & tana.jones (???) \\ 

2 & mark.taylor (EMP) & sara.shackleton (???) & sara.shackleton (???) & jeff.dasovich (EMP) \\ 

3 & sara.shackleton (???) & mark.taylor (EMP) & jeff.dasovich (EMP) & sara.shackleton (???) \\ 

4 & carol.clair (LAW) & carol.clair (LAW) & mark.taylor (EMP) & bill.williams (???) \\ 

5 & jeff.dasovich (EMP) & marie.heard (???) & chris.germany (EMP) & mike.grigsby (MAN) \\ 

6 & eric.bass (TRA) & jeff.dasovich (EMP) & eric.bass (TRA) & chris.germany (EMP) \\ 

7 & steven.kean (VP) & mark.haedicke (MD) & carol.clair (LAW) & mark.taylor (EMP) \\ 

8 & mark.haedicke (MD) & d..steffes (VP) & susan.scott (???) & eric.bass (TRA) \\ 

9 & elizabeth.sager (EMP) & elizabeth.sager (EMP) & scott.neal (VP) & john.arnold (VP) \\ 

10 & mary.hain (LAW) & eric.bass (TRA) & drew.fossum (VP) & scott.neal (VP) \\ 

11 & richard.sanders (VP) & steven.kean (VP) & mike.grigsby (MAN) & phillip.love (???) \\ 

12 & phillip.allen (???) & louise.kitchen (PRE) & david.delainey (CEO) & phillip.allen (???) \\ 

13 & susan.scott (???) & richard.sanders (VP) & phillip.allen (???) & susan.scott (???) \\ 

14 & bill.williams (???) & bill.williams (???) & sally.beck (EMP) & debra.perlingiere (???) \\ 

15 & chris.germany (EMP) & mike.grigsby (MAN) & debra.perlingiere (???) & kimberly.watson (???) \\ 

16 & mike.grigsby (MAN) & mary.hain (LAW) & john.arnold (VP) & steven.kean (VP) \\ 

17 & sally.beck (EMP) & kim.ward (???) & bill.williams (???) & louise.kitchen (PRE) \\ 

18 & drew.fossum (VP) & phillip.love (???) & elizabeth.sager (EMP) & sally.beck (EMP) \\ 

19 & david.delainey (CEO) & chris.germany (EMP) & richard.sanders (VP) & david.delainey (CEO) \\ 

20 & matthew.lenhart (EMP) & gerald.nemec (???) & gerald.nemec (???) & carol.clair (LAW) \\ 

21 & gerald.nemec (???) & phillip.allen (???) & mark.haedicke (MD) & mary.hain (LAW) \\ 

22 & phillip.love (???) & matthew.lenhart (EMP) & matthew.lenhart (EMP) & drew.fossum (VP) \\ 

23 & scott.neal (VP) & kay.mann (EMP) & phillip.love (???) & d..steffes (VP) \\ 

24 & d..steffes (VP) & sally.beck (EMP) & steven.kean (VP) & gerald.nemec (???) \\ 

25 & kay.mann (EMP) & john.arnold (VP) & mary.hain (LAW) & matthew.lenhart (EMP) \\ 

26 & debra.perlingiere (???) & david.delainey (CEO) & darron.giron (EMP) & darron.giron (EMP) \\ 

27 & john.arnold (VP) & susan.scott (???) & mike.mcconnell (???) & john.lavorato (CEO) \\ 

28 & darron.giron (EMP) & debra.perlingiere (???) & kay.mann (EMP) & kay.mann (EMP) \\ 

29 & jane.tholt (VP) & scott.neal (VP) & kate.symes (EMP) & richard.sanders (VP) \\ 

30 & mike.mcconnell (???) & drew.fossum (VP) & john.lavorato (CEO) & marie.heard (???) \\ 

31 & john.lavorato (CEO) & darron.giron (EMP) & dan.hyvl (EMP) & kate.symes (EMP) \\ 

32 & kimberly.watson (???) & barry.tycholiz (VP) & jane.tholt (VP) & elizabeth.sager (EMP) \\ 

33 & lynn.blair (???) & kimberly.watson (???) & kimberly.watson (???) & lynn.blair (???) \\ 

34 & louise.kitchen (PRE) & john.lavorato (CEO) & d..steffes (VP) & mark.haedicke (MD) \\ 

35 & dan.hyvl (EMP) & jane.tholt (VP) & jeffrey.shankman (PRE) & errol.mclaughlin (EMP) \\ 

36 & kim.ward (???) & dan.hyvl (EMP) & errol.mclaughlin (EMP) & mike.mcconnell (???) \\ 

37 & errol.mclaughlin (EMP) & mike.mcconnell (???) & louise.kitchen (PRE) & kevin.presto (VP) \\ 

38 & marie.heard (???) & kevin.presto (VP) & hunter.shively (VP) & kim.ward (???) \\ 

39 & jeffrey.shankman (PRE) & errol.mclaughlin (EMP) & marie.heard (???) & dan.hyvl (EMP) \\ 

40 & kate.symes (EMP) & lynn.blair (???) & lynn.blair (???) & michelle.lokay (EMP) \\ 

41 & barry.tycholiz (VP) & michelle.cash (???) & michelle.lokay (EMP) & rod.hayslett (VP) \\ 

42 & kevin.presto (VP) & kam.keiser (EMP) & kim.ward (???) & jane.tholt (VP) \\ 

43 & tracy.geaccone (EMP) & rod.hayslett (VP) & rob.gay (???) & tracy.geaccone (EMP) \\ 

44 & hunter.shively (VP) & stacy.dickson (EMP) & kevin.presto (VP) & barry.tycholiz (VP) \\ 

45 & darrell.schoolcraft (???) & michelle.lokay (EMP) & chris.dorland (EMP) & mark.whitt (???) \\ 

46 & michelle.lokay (EMP) & kenneth.lay (CEO) & fletcher.sturm (VP) & john.forney (MAN) \\ 

47 & rod.hayslett (VP) & tracy.geaccone (EMP) & robin.rodrigue (???) & chris.dorland (EMP) \\ 

48 & rob.gay (???) & jeffrey.shankman (PRE) & tracy.geaccone (EMP) & jeffrey.shankman (PRE) \\ 

49 & robin.rodrigue (???) & fletcher.sturm (VP) & rod.hayslett (VP) & darrell.schoolcraft (???) \\ 

50 & robert.badeer (DIR) & kate.symes (EMP) & andrea.ring (???) & kam.keiser (EMP) \\ 

51 & tori.kuykendall (TRA) & susan.bailey (???) & barry.tycholiz (VP) & hunter.shively (VP) \\ 

52 & greg.whalley (VP) & mark.whitt (???) & greg.whalley (VP) & kenneth.lay (CEO) \\ 

53 & kenneth.lay (CEO) & tori.kuykendall (TRA) & tori.kuykendall (TRA) & bill.rapp (???) \\ 

54 & fletcher.sturm (VP) & hunter.shively (VP) & john.forney (MAN) & lindy.donoho (EMP) \\ 

55 & chris.dorland (EMP) & martin.cuilla (MAN) & michelle.cash (???) & fletcher.sturm (VP) \\ 

56 & peter.keavey (EMP) & james.derrick (LAW) & peter.keavey (EMP) & shelley.corman (VP) \\ 

57 & bill.rapp (???) & jeffrey.hodge (MD) & mark.guzman (TRA) & martin.cuilla (MAN) \\ 

58 & michelle.cash (???) & jeff.skilling (CEO) & darrell.schoolcraft (???) & tori.kuykendall (TRA) \\ 

59 & daren.farmer (MAN) & andy.zipper (VP) & kenneth.lay (CEO) & kevin.hyatt (DIR) \\ 

60 & lindy.donoho (EMP) & darrell.schoolcraft (???) & larry.may (DIR) & andrea.ring (???) \\ 

61 & mark.whitt (???) & chris.dorland (EMP) & daren.farmer (MAN) & rob.gay (???) \\ 

62 & larry.may (DIR) & bill.rapp (???) & martin.cuilla (MAN) & andy.zipper (VP) \\ 

63 & benjamin.rogers (???) & greg.whalley (VP) & mark.whitt (???) & greg.whalley (VP) \\ 

64 & john.forney (MAN) & dutch.quigley (???) & jeff.skilling (CEO) & dutch.quigley (???) \\ 

65 & martin.cuilla (MAN) & lindy.donoho (EMP) & rick.buy (MAN) & jeff.skilling (CEO) \\ 

66 & andy.zipper (VP) & shelley.corman (VP) & james.derrick (LAW) & rick.buy (MAN) \\ 

67 & shelley.corman (VP) & patrice.mims (???) & patrice.mims (???) & t..lucci (EMP) \\ 

68 & jeff.skilling (CEO) & monique.sanchez (???) & dutch.quigley (???) & robin.rodrigue (???) \\ 

69 & monique.sanchez (???) & peter.keavey (EMP) & shelley.corman (VP) & james.derrick (LAW) \\ 

70 & kam.keiser (EMP) & rick.buy (MAN) & benjamin.rogers (???) & jonathan.mckay (DIR) \\ 

71 & dutch.quigley (???) & rob.gay (???) & lindy.donoho (EMP) & jim.schwieger (TRA) \\ 

72 & mark.guzman (TRA) & robin.rodrigue (???) & bill.rapp (???) & larry.may (DIR) \\ 

73 & rick.buy (MAN) & thomas.martin (VP) & kam.keiser (EMP) & monique.sanchez (???) \\ 

74 & kevin.hyatt (DIR) & kevin.hyatt (DIR) & mike.carson (EMP) & michelle.cash (???) \\ 

75 & james.derrick (LAW) & larry.may (DIR) & dana.davis (???) & mark.guzman (TRA) \\ 

76 & andrea.ring (???) & joe.parks (???) & andy.zipper (VP) & thomas.martin (VP) \\ 

77 & stacy.dickson (EMP) & john.forney (MAN) & monique.sanchez (???) & teb.lokey (MAN) \\ 

78 & patrice.mims (???) & jim.schwieger (TRA) & kevin.ruscitti (TRA) & patrice.mims (???) \\ 

79 & jim.schwieger (TRA) & john.zufferli (EMP) & judy.hernandez (???) & diana.scholtes (TRA) \\ 

80 & jonathan.mckay (DIR) & daren.farmer (MAN) & jim.schwieger (TRA) & peter.keavey (EMP) \\ 

81 & kevin.ruscitti (TRA) & t..lucci (EMP) & stacy.dickson (EMP) & john.zufferli (EMP) \\ 

82 & t..lucci (EMP) & jonathan.mckay (DIR) & larry.campbell (???) & daren.farmer (MAN) \\ 

83 & sandra.brawner (DIR) & richard.ring (EMP) & kevin.hyatt (DIR) & stacy.dickson (EMP) \\ 

84 & geir.solberg (EMP) & andrea.ring (???) & t..lucci (EMP) & sandra.brawner (DIR) \\ 

85 & jeffrey.hodge (MD) & judy.townsend (EMP) & jonathan.mckay (DIR) & matt.smith (???) \\ 

86 & geoff.storey (DIR) & robert.badeer (DIR) & thomas.martin (VP) & danny.mccarty (VP) \\ 

87 & thomas.martin (VP) & teb.lokey (MAN) & sandra.brawner (DIR) & cara.semperger (EMP) \\ 

88 & teb.lokey (MAN) & mark.guzman (TRA) & jeffrey.hodge (MD) & larry.campbell (???) \\ 

89 & matt.smith (???) & doug.gilbert-smith (MAN) & judy.townsend (EMP) & dana.davis (???) \\ 

90 & john.zufferli (EMP) & diana.scholtes (TRA) & matt.smith (???) & benjamin.rogers (???) \\ 

91 & judy.townsend (EMP) & geoff.storey (DIR) & john.zufferli (EMP) & jeffrey.hodge (MD) \\ 

92 & danny.mccarty (VP) & danny.mccarty (VP) & jason.williams (???) & ryan.slinger (TRA) \\ 

93 & diana.scholtes (TRA) & sandra.brawner (DIR) & diana.scholtes (TRA) & joe.parks (???) \\ 

94 & jay.reitmeyer (EMP) & jay.reitmeyer (EMP) & teb.lokey (MAN) & sean.crandall (DIR) \\ 

95 & holden.salisbury (EMP) & charles.weldon (???) & sean.crandall (DIR) & jason.williams (???) \\ 

96 & frank.ermis (DIR) & matt.smith (???) & paul.thomas (???) & paul.thomas (???) \\ 

97 & ryan.slinger (TRA) & benjamin.rogers (???) & charles.weldon (???) & jay.reitmeyer (EMP) \\ 

98 & larry.campbell (???) & ryan.slinger (TRA) & danny.mccarty (VP) & geoff.storey (DIR) \\ 

99 & joe.parks (???) & cara.semperger (EMP) & ryan.slinger (TRA) & mike.carson (EMP) \\ 

100 & dana.davis (???) & geir.solberg (EMP) & geir.solberg (EMP) & geir.solberg (EMP) \\ 

101 & sean.crandall (DIR) & sean.crandall (DIR) & geoff.storey (DIR) & kevin.ruscitti (TRA) \\ 

102 & cara.semperger (EMP) & kevin.ruscitti (TRA) & susan.pereira (EMP) & judy.hernandez (???) \\ 

103 & mike.carson (EMP) & jason.wolfe (???) & frank.ermis (DIR) & charles.weldon (???) \\ 

104 & paul.y'barbo (???) & scott.hendrickson (???) & robert.badeer (DIR) & judy.townsend (EMP) \\ 

105 & andrew.lewis (DIR) & holden.salisbury (EMP) & joe.parks (???) & holden.salisbury (EMP) \\ 

106 & charles.weldon (???) & keith.holst (DIR) & jay.reitmeyer (EMP) & theresa.staab (EMP) \\ 

107 & jason.williams (???) & susan.pereira (EMP) & cara.semperger (EMP) & paul.y'barbo (???) \\ 

108 & paul.thomas (???) & frank.ermis (DIR) & holden.salisbury (EMP) & vladi.pimenov (???) \\ 

109 & jason.wolfe (???) & albert.meyers (EMP) & jeff.king (MAN) & don.baughman (TRA) \\ 

110 & susan.pereira (EMP) & dana.davis (???) & paul.y'barbo (???) & jeff.king (MAN) \\ 

111 & mike.swerzbin (TRA) & paul.y'barbo (???) & theresa.staab (EMP) & susan.pereira (EMP) \\ 

112 & judy.hernandez (???) & larry.campbell (???) & andrew.lewis (DIR) & doug.gilbert-smith (MAN) \\ 

113 & theresa.staab (EMP) & mike.swerzbin (TRA) & scott.hendrickson (???) & jason.wolfe (???) \\ 

114 & scott.hendrickson (???) & theresa.staab (EMP) & jason.wolfe (???) & harry.arora (VP) \\ 

115 & mike.maggi (DIR) & mike.carson (EMP) & vince.kaminski (MAN) & frank.ermis (DIR) \\ 

116 & keith.holst (DIR) & don.baughman (TRA) & don.baughman (TRA) & john.griffith (MD) \\ 

117 & jeff.king (MAN) & jason.williams (???) & tom.donohoe (???) & eric.saibi (TRA) \\ 

118 & vladi.pimenov (???) & john.griffith (MD) & vladi.pimenov (???) & mike.swerzbin (TRA) \\ 

119 & don.baughman (TRA) & paul.thomas (???) & mike.maggi (DIR) & scott.hendrickson (???) \\ 

120 & richard.shapiro (VP) & vladi.pimenov (???) & mike.swerzbin (TRA) & keith.holst (DIR) \\ 

121 & vince.kaminski (MAN) & judy.hernandez (???) & harry.arora (VP) & richard.ring (EMP) \\ 

122 & harry.arora (VP) & mike.maggi (DIR) & eric.saibi (TRA) & vince.kaminski (MAN) \\ 

123 & susan.bailey (???) & pam.butler (???) & john.griffith (MD) & susan.bailey (???) \\ 

124 & doug.gilbert-smith (MAN) & jeff.king (MAN) & keith.holst (DIR) & mike.maggi (DIR) \\ 

125 & john.griffith (MD) & andrew.lewis (DIR) & doug.gilbert-smith (MAN) & robert.badeer (DIR) \\ 

126 & eric.saibi (TRA) & vince.kaminski (MAN) & susan.bailey (???) & tom.donohoe (???) \\ 

127 & richard.ring (EMP) & harry.arora (VP) & cooper.richey (MAN) & albert.meyers (EMP) \\ 

128 & tom.donohoe (???) & eric.saibi (TRA) & richard.ring (EMP) & clint.dean (TRA) \\ 

129 & clint.dean (TRA) & richard.shapiro (VP) & joe.stepenovitch (VP) & andrew.lewis (DIR) \\ 

130 & albert.meyers (EMP) & tom.donohoe (???) & clint.dean (TRA) & cooper.richey (MAN) \\ 

131 & cooper.richey (MAN) & clint.dean (TRA) & joe.quenet (TRA) & joe.stepenovitch (VP) \\ 

132 & pam.butler (???) & cooper.richey (MAN) & albert.meyers (EMP) & pam.butler (???) \\ 

133 & joe.stepenovitch (VP) & joe.stepenovitch (VP) & pam.butler (???) & steven.merris (???) \\ 

134 & joe.quenet (TRA) & stephanie.panus (EMP) & richard.shapiro (VP) & richard.shapiro (VP) \\ 

135 & stephanie.panus (EMP) & joe.quenet (TRA) & steven.merris (???) & monika.causholli (EMP) \\ 

136 & stanley.horton (PRE) & brad.mckay (EMP) & phillip.platter (EMP) & mark.fisher (???) \\ 

137 & steven.merris (???) & stanley.horton (PRE) & stanley.horton (PRE) & phillip.platter (EMP) \\ 

138 & brad.mckay (EMP) & phillip.platter (EMP) & mark.fisher (???) & stephanie.panus (EMP) \\ 

139 & phillip.platter (EMP) & steven.merris (???) & monika.causholli (EMP) & stanley.horton (PRE) \\ 

140 & monika.causholli (EMP) & monika.causholli (EMP) & brad.mckay (EMP) & joe.quenet (TRA) \\ 

141 & mark.fisher (???) & mark.fisher (???) & stephanie.panus (EMP) & brad.mckay (EMP) \\ 
\end{longtable*} 
\end{appendices}
\end{document}